%% file: ms.tex
\documentclass[12pt,preprint]{aastex}

\newcommand\be{\begin{equation}}
\newcommand\en{\end{equation}}

\shorttitle{GM Cep}
\shortauthors{Sicilia-Aguilar et al.}

\begin{document}

\title{The Rapid Outbursting Star GM Cep: An EX-or in Tr 37?}

\author{Aurora Sicilia-Aguilar\altaffilmark{1}, Bruno Mer\'{\i}n\altaffilmark{2}, Felix Hormuth\altaffilmark{1}, P\'{e}ter \'{A}brah\'{a}m\altaffilmark{3},}
\author{Thomas Henning\altaffilmark{1},  M\'{a}ria Kun\altaffilmark{3}, Nimesh Patel\altaffilmark{4}, Attila Juh\'{a}sz\altaffilmark{1}, Wolfgang Brandner\altaffilmark{1},}
\author{Lee W. Hartmann\altaffilmark{5}, Szil\'{a}rd Csizmadia\altaffilmark{3}, Attila Mo\'{o}r\altaffilmark{3}}

\altaffiltext{1}{Max-Planck-Institut f\"{u}r Astronomie, K\"{o}nigstuhl 17, 69117 Heidelberg, Germany}
\altaffiltext{2}{Research and Scientific Support Department, European Space Agency, ESTEC, Keplerlaan 1, 2200 AG Noordwijk, The Netherlands}
\altaffiltext{3}{Konkoly Observatory, H-1525 Budapest, P.O.Box 67, Hungary}
\altaffiltext{4}{Center for Astrophysics, 60 Garden Street, Cambridge, MA 02138}
\altaffiltext{5}{University of Michigan, 830 Dennison 500 Church St., Ann Arbor, MI 48109}

\email{sicilia@mpia.de}

\begin{abstract}

We present optical, IR and millimeter observations of the solar-type star 13-277, 
also known as GM Cep, in the 4 Myr-old cluster Tr 37. GM Cep experiences rapid
magnitude variations of more than 2 mag at optical wavelengths. We explore the 
causes of the variability, which seem to be dominated by strong increases in the 
accretion, being similar to EX-or episodes. The star shows high, variable accretion 
rates (up to $\sim$10$^{-6}$ M$_\odot$/yr), signs of powerful winds, 
and it is a very fast rotator (V$sini \sim$43 km/s). Its strong mid-IR excesses reveal
a very flared disk and/or a remnant envelope, most likely out of hydrostatic 
equilibrium. The 1.3 millimeter fluxes suggest a relatively  massive disk
(M$_D \sim$0.1 M$_\odot$). Nevertheless, the millimeter mass is not enough to sustain
increased accretion episodes over large timescales, unless the mass is underestimated 
due to significant grain growth. We finally explore the possibility of GM Cep having
a binary companion, which could trigger disk instabilities producing the enhanced 
accretion episodes.

\end{abstract}


\section{Introduction \label{intro}}

The open cluster Tr 37 is one of the best studied intermediate-aged 
young star formation regions. Aged $\sim$ 4 Myr (Sicilia-Aguilar et al.
2004, 2005; hereafter Paper I, Paper II), and located at 900 pc distance 
(Contreras et al. 2002), it contains a rich population of T Tauri stars 
(TTS; $\sim$180 members with spectral types G to M2) of which about 48\% 
still show IR excesses consistent with protoplanetary disks at different 
evolutionary stages (Sicilia-Aguilar et al. 2006a, b; hereafter Paper III, 
Paper IV). While most of the stars in Tr 37 show important evidence of
disk evolution, with lower accretion rates and near-IR excesses than in 
younger regions, a few objects still display characteristics of much 
younger systems. The most remarkable one is the solar-type star 13-277, 
also called GM Cep (Morgenroth 1939). Because of its spatial location within 
the cluster, it belongs most likely to the Tr 37 main population, with 
average ages $\sim$4 Myr ($\sim$85\% of the population is older than 2 Myr, 
and $\sim$95\% is older than 1 Myr), rather than to the young population 
associated to the Tr 37 globule, aged $\sim$1 Myr (Paper II). Its continuum 
spectrum and the strong and broad H$\alpha$ emission suggested an accretion 
rate \.{M}$\sim$3 10$^{-7}$ M$_\odot$/yr, about 2 orders of magnitude over 
the median accretion rate of TTS in Tr 37 (Paper IV). Its bolometric 
luminosity  (L$\sim$26 L$_\odot$ in 2000) is about 1 order of magnitude higher 
than other stars with similar late G-early K spectral type. Finally, we notice 
an unusually high mid-IR flux for a solar-type star: its MIPS flux at 70 $\mu$m 
is comparable to that of the Tr 37 Herbig Be star MVA-426, being one of the only 
four cluster members detected at this wavelength (Paper III). 

All observations suggest that GM Cep is a variable star of the EX-or 
type, with an unstable disk and variable accretion rate, which is
remarkable in an ``old'' cluster like Tr 37, where disk evolution seems
ubiquitous. FU-ors and EX-or objects have been suggested to be either normal stages 
within the very early TTS evolution (see Hartmann \& Kenyon 1996 for a review) or 
a special type of young object, probably binary (Herbig 2003 and references therein). 
Here we present the results of our photometry monitoring campaign during 
2006-2007\footnote{Based on observations collected at the German-Spanish Astronomical 
Center, Calar Alto, jointly operated by the Max-Planck-Institut f\"{u}r Astronomie 
Heidelberg and the  Instituto de Astrof\'{\i}sica de Andaluc\'{\i}a (CSIC)}, lucky 
imaging, millimeter continuum, and $^{12}$CO data, together with optical spectra 
(high- and low-resolution), and a compilation of the data available from the 
literature. The multiwavelength data are described in Section \ref{observations}. In 
Section \ref{analysis} we explore the characteristics of the star: optical and IR 
variability, spectral type, accretion, winds, disk mass, and potential companions. 
Section \ref{outburst} discusses the possible causes of variability and the comparison 
with similar stars, and in Section \ref{conclu} we summarize our results.

\section{Observations and Data Reduction \label{observations}}

\subsection{Data from the Literature \label{data-literature}}

The photometric data and epochs available for GM Cep are given in Table \ref{obs-table}. 
The first reference to GM Cep, Morgenroth (1939), classifies it as a ``long-period 
variable'' with visual magnitudes 13.5-15.5, although no epoch nor period are mentioned. 
Suyarkova (1975) studied the light variations over several months,
detecting rapid variations (days to weeks) between 14.2 and 16.4 mag, mixed 
with stability periods up to $\sim$100 days at both the lower and the upper 
magnitude. Kun (1986) lists the star among the variables in Cep OB2, providing V 
photometry for three epochs. The databases Vizier, Simbad,
and SuperCosmos list optical photometric data from the USNO-A2.0, USNO-B1.0, 
GSC2.2, and SuperCosmos catalogs, plus IR data from IRAS and MSX6C
(Egan et al. 2003). There are two mentions of GM Cep in the amateur database of 
the American Association of Variable Star Observers (AAVSO), but they do not add 
any extra information to our data. Our previous studies of Tr 37 include 
optical photometry taken with the 1.2m telescope at the Fred Lawrence Whipple 
Observatory (FLWO) between 2000 and 2004 (Paper I, II), low resolution optical 
spectroscopy taken with the FAST spectrograph on the 1.5m telescope at the FLWO 
(Paper I), high-resolution optical spectroscopy from Hectochelle/MMT at the FLWO 
(Paper IV), and IRAC and MIPS data from our Spitzer study (Paper III). We also 
include the near-IR data from 2MASS, presented in Paper I. All these 
data confirmed the variability of GM Cep and the presence of strong IR excesses 
from a very luminous disk. 

In order to compare the data from the different studies in Table \ref{obs-table}, 
we transformed the photographic and CCD data from different filters into the 
UBVR$_c$I$_c$ system, which is the closest match to the data prior to $\sim$1986. 
The photographic R63F and R59F red data are comparable to the modern R$_c$ filter. 
R$_c$-R63F=0.096 and 0.081 mag, for giants and dwarfs, respectively, and 
R$_c$-R59F=-0.031 (Bessell 1986). Following Blair \& Gilmore (1982), R$_c$=R63F if 
R$_c$-I$_c <$0.9 mag (for GM Cep, R$_c$-I$_c$ $\sim$0.8 mag). The GG395-B$_j$ 
photographic passband is similar to the Johnsons B filter, although the zero point 
is different (Bessell et al. 1986). The errors involved in these conversions are 
less than 0.1 mag, below the typical errors from the photographic plates 
and well below the observed variations of GM Cep. For the modern observations, 
when contemporary V magnitude is available, R$_J$ and I$_J$ can be transformed 
into R$_c$ and I$_c$ for comparison (Bessel 1979; Fernie 1983) with errors under 
0.01 mag for R, and around 0.1-0.2 mag for I.

\subsection{Optical Photometry \label{data-optical}}

The first set of optical observations was taken with the KING 70 cm telescope of 
the Max-Planck-Institut f\"{u}r Astronomie, at the K\"{o}nigstuhl in Heidelberg
(see Table \ref{obs-table}). The telescope has a field of view of 
$\sim$18'x18', and is equipped with a Loral CCD L3-W17 detector and a set of 
standard UBVR$_J$I$_J$ filters. The filters have very small color transform 
coefficients, and the second order extinction is almost zero, so relative 
photometry is feasible. Exposure times ranged from 30s to 3 x 120s depending on 
filter and weather conditions. We also obtained VR$_J$I$_J$ photometry  
(Table \ref{obs-table})with the CAFOS imager on the 2.2m telescope in Calar 
Alto (Director's Discretionary Time, DDT), which covers a field of view of diameter 16'. 
Taking 3 x 10 s or 3 x 30s exposures, we reach similar magnitude and area coverage 
as the K\"{o}nigstuhl observations. 

The data were reduced following standard routines within the IRAF\footnote{IRAF 
is distributed by the National Optical Astronomy Observatories,
which are operated by the Association of Universities for Research
in Astronomy, Inc., under cooperative agreement with the National
Science Foundation.} tasks \textit{noao.imred.ccdred} to do the bias and flat 
field corrections. Aperture photometry was performed within IRAF 
\textit{noao.digiphot.apphot}, and the data in each band were calibrated via 
relative photometry, comparing with our previous observations from 
the FLWO (Paper I). Depending on the night and instrument, $\sim$150-800 stars 
were used for calibrating VRI, and $\sim$20-200 for B and U. Since the FLWO 
observations used the UVR$_c$I$_c$ filters, the R$_c$ and I$_c$ bands were 
transformed into the Johnsons system (Fernie 1983; Bessel 1979). Taking
into account the large number of stars involved in the calibration, the errors 
derived from band transformation are minimal and the final error is dominated by 
the signal to noise, being typically less than 0.01 mag. Due to the lack of 
calibrated B data in the FLWO survey, we used the B photometry from the literature 
for the relative photometry. About 150 stars with B band photometry are listed by
SIMBAD, mostly bright objects from the Marshall \& Van Altena (1987) catalog.
Only 15 of these stars are not saturated in the K\"{o}nigstuhl frames and could
be used for calibration. Whereas this is insufficient for 
studying colors, the relative changes between different nights are unaffected. 

\subsection{Optical Spectroscopy: HIRES/Keck and CAFOS/2.2m\label{data-spec}}

GM Cep was observed with the high-resolution spectrograph HIRES mounted on Keck I on 
2001 June 30. The spectra were taken before the HIRES upgrades in 2004, using the 
rg610 filter and a 1.148'' slit, which provides a wavelength coverage between 6350 
and 8750 \AA\ , with some gaps in between orders, and a resolution R$\sim$ 34,000.
The resolution is similar to that of Hectochelle/MMT (Sicilia-Aguilar 
et al. 2005a; Paper IV), but the wavelength coverage is much larger, including
the H$\alpha$ and Li I 6708\AA\ regions, several forbidden lines ([O I] at 6368 \AA, 
[S II] at 6717 and 6731 \AA, [N II] at 6548 and 6583 \AA), the Ca II IR lines at 8498 
and 8662 \AA, and numerous photospheric absorption lines. The exposure time was 3 x 900 s. 
The data was reduced using the MAKEE Keck Observatory HIRES data reduction software, 
developed by T. Barlow. The wavelength calibration was done with a ThAr lamp, being 
accurate up to 1-2 km/s. No flux calibration was required.

A low-resolution spectrum was taken with the CAFOS camera in the long-slit spectroscopic 
mode on the 2.2m telescope  as a part of our Calar Alto DDT program. We used the 
B-100, G-100, and R-100 grisms, resulting in a wavelength coverage from 3200 to 9000 \AA\ 
with no gaps and some overlap, and a resolution of 2 \AA\ per pixel (Figure \ref{lowrespec-fig}). 
The spectra were reduced using the standard IRAF tasks (\textit{noao.imred.ccdred} 
and \textit{specred.twodspec}), and a wavelength solution was derived using a HgHeRb lamp. 
A total of 3 x 600 s exposures were taken with each grism, in order to achieve a good 
cosmic ray removal. In addition, we obtained two CAFOS acquisition images in 
V and R$_J$, with exposure times of 60s. These two images, although lower in quality and 
with a smaller field of view than the standard CAFOS imaging, were reduced and calibrated 
using the techniques from Section \ref{data-optical}, to provide simultaneous photometry.

\subsection{Lucky Imaging with AstraLux \label{data-astra}}

\begin{figure}
\plotone{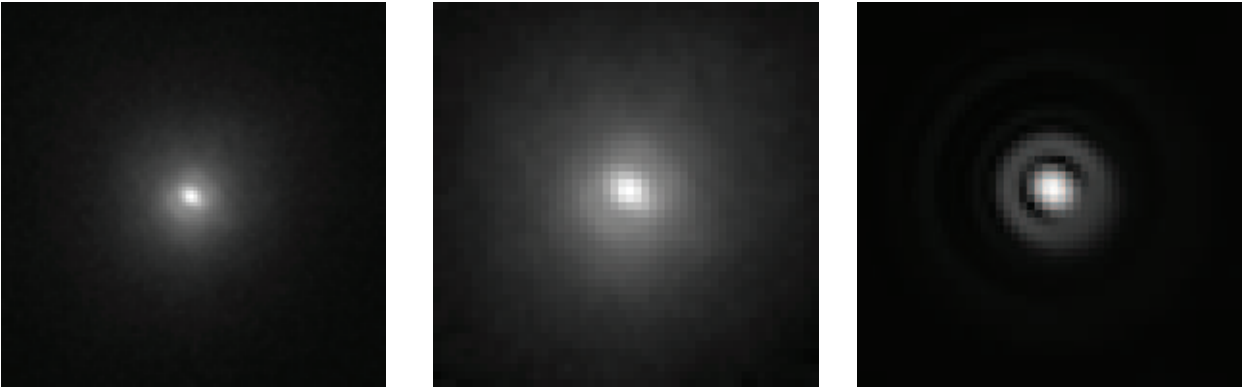}
\caption{AstraLux observations of GM Cep in the SDSS z' filter. The left image has a 
field of view of 2\arcsec$\times$2\arcsec\ and was generated from the best 
2\% of 15,000 single frames. The effective integration time is 7.5\,s at a 
single frame exposure time of 25\,ms. The middle image shows the inner 
0\farcs9$\times$0\farcs9. The right image is the theoretical PSF with same 
pixel scale, taking atmospheric dispersion effects into account. 
All images are square root scaled up to saturation. \label{astralux-fig}}
\end{figure}

In order to look for wide- and intermediate-distance companions, we observed
GM Cep with the high spatial resolution Lucky Imaging camera AstraLux, operating 
at the 2.2m telescope in Calar Alto. AstraLux is a high-speed camera
with an electron multiplying CCD, which allows full frame rates of up to 
34\,Hz with virtually zero readout noise. It has a field of view  of
24''$\times$24'' at a pixel scale of 47\,mas/px.
The AstraLux data consist typically of several thousands single frames with 
integration times between 15 and 100\,ms. Software post-processing selects 
the frames which are least affected by distortions due to atmospheric turbulence, 
by measuring the Strehl ratio in each single frame. The 
high-quality frames ($\sim$ 1--5\% of all images) are combined to a final result 
with improved angular resolution and Strehl ratio (Lucky Imaging technique; 
Tubbs et al. 2002; Law et al. 2006). Under good seeing conditions and with the 
SDSS z' filter, AstraLux provides spatial resolution better than 100\,mas and 
Strehl ratios of up to 25\%, depending on the magnitude. Observations of GM Cep 
were conducted in July and November 2006, and May and June 2007 
(see Tables \ref{obs-table} and \ref{astralux-table} for details). 
Though the November 2006 data is slightly affected by atmospheric  
dispersion, leading to an elongated PSF, the superb seeing of 0.6''
still resulted in a Strehl ratio of 14\%. Figure \ref{astralux-fig} shows
the November 2006 data together with a simulated theoretical PSF,
including atmospheric dispersion effects. We present a detailed analysis of 
these data in Section \ref{companion}. 

\subsection{JHK Photometry \label{data-JHK}}

Near-IR photometry was obtained with the IR camera CAIN-2 at the 
1.52m Carlos S\'{a}nchez telescope in the Teide Observatory (Canary Islands)
during 5 nights in early June 2007 (see Table \ref{obs-table}). Observations 
in the J H and Ks bands consist of 5 dithered exposures around GM Cep to ensure 
proper sky image subtraction. For each dither point we took several frames 
using short exposure time to avoid saturation, which were later averaged to 
obtain the final reduced image. The minimum exposure time was 1~s for individual 
exposures and the total integration time per filter was about 6 min. The data were 
reduced using the package \textit{caindr} developed within IRAF by 
J.A. Acosta-Pulido. The processing included sky subtraction, flatfielding 
and combination of the dither positions. Aperture photometry was done, and the 
relative photometry calibration was based on 5 to 13 nearby IR sources taken from 
the 2MASS catalog. The errors derived from this procedure are dominated by the 
background and CCD read noise, ranging from 0.1 to 0.01 mag depending on weather 
conditions. The star remained stable during the whole run (see Table \ref{JHK-table}), 
taking into account the errors, and its magnitude was similar to the  
2MASS 2001 data. Similar stability over a few days had been observed in the optical data
taken in September 2000. Simultaneous optical observations cover only 1 of the nights, 
during which the star was near maximum.

\subsection{IRAM 1.3 Millimeter Continuum \label{data-mm}}
 
We observed GM Cep at the IRAM 30m telescope in Pico Veleta using the 37 channel 
bolometer, MAMBO-1. The observations were done during night time, at 4:18h UT on 
June 2nd 2006, and 2:58h on June 3rd. The opacity at zenith was Tau=0.4.
Each observation consisted of a 20 min integration standard \textit{onoff}, 
with 10 min integration on source. The wobbler throw for the \textit{off} 
position was 32'' and 50'', respectively, so the \textit{off} measurement falls 
in different locations, to avoid the potential contamination from other sources. 
The data were reduced using the MOPSIC software, developed by R. Zylka for the 30m 
bolometer data. The pointing and flux calibration were done with the sources 
LK H$\alpha$ 234, Cep A, and NGC 7538 from the IRAM Pool catalog, and the planet 
Uranus. The final flux at 1.3mm is 13.9 $\pm$ 1.0 mJy. The 1.3 millimeter continuum 
observations are analyzed in Section \ref{diskmodels}. 
 
\subsection{IRAM $^{12}$CO(1-0) and $^{12}$CO(2-1) Observations \label{data-co}}
 
We also observed GM Cep with the heterodyne receivers at the 30m IRAM 
telescope. The observations were performed under fair weather conditions 
on June 3rd (UT 7:00-8:15), June 4th (UT 7:30-9:20), and June 5th 
(UT 8:00-10:00). We used the standard \textit{onoff} mode with wobbler shift, 
that results in better baselines, with wobbler throws from 70'' to 120''. 
We used the highest resolution mode available (2 km/s), 
and the backends A100, B100 (for frequencies around 115 GHz)
and A230, B230 (for frequencies around 230 GHz). The instrument allows to 
observe two frequencies simultaneously, which we tuned to the $^{12}$CO(1-0) 
and $^{12}$CO(2-1) transitions. Since the LRS velocity of the source (considering 
the average cluster CZ=-15.0$\pm$3.6 km/s) at the time of the observation
was around 0 km/s, the receivers were tuned to 115.271-115.275 GHz 
and 230.538-230.543 GHz, varying the central frequencies during different 
integrations to avoid that the line falls always on the same place 
on the detector. The total integration time was 4 h, with 
2h on source. For pointing and calibration, we observed the standard
sources J2253+161, NGC 7027, and Cep A, from the IRAM Pool catalog.
 
\begin{figure}
\plottwo{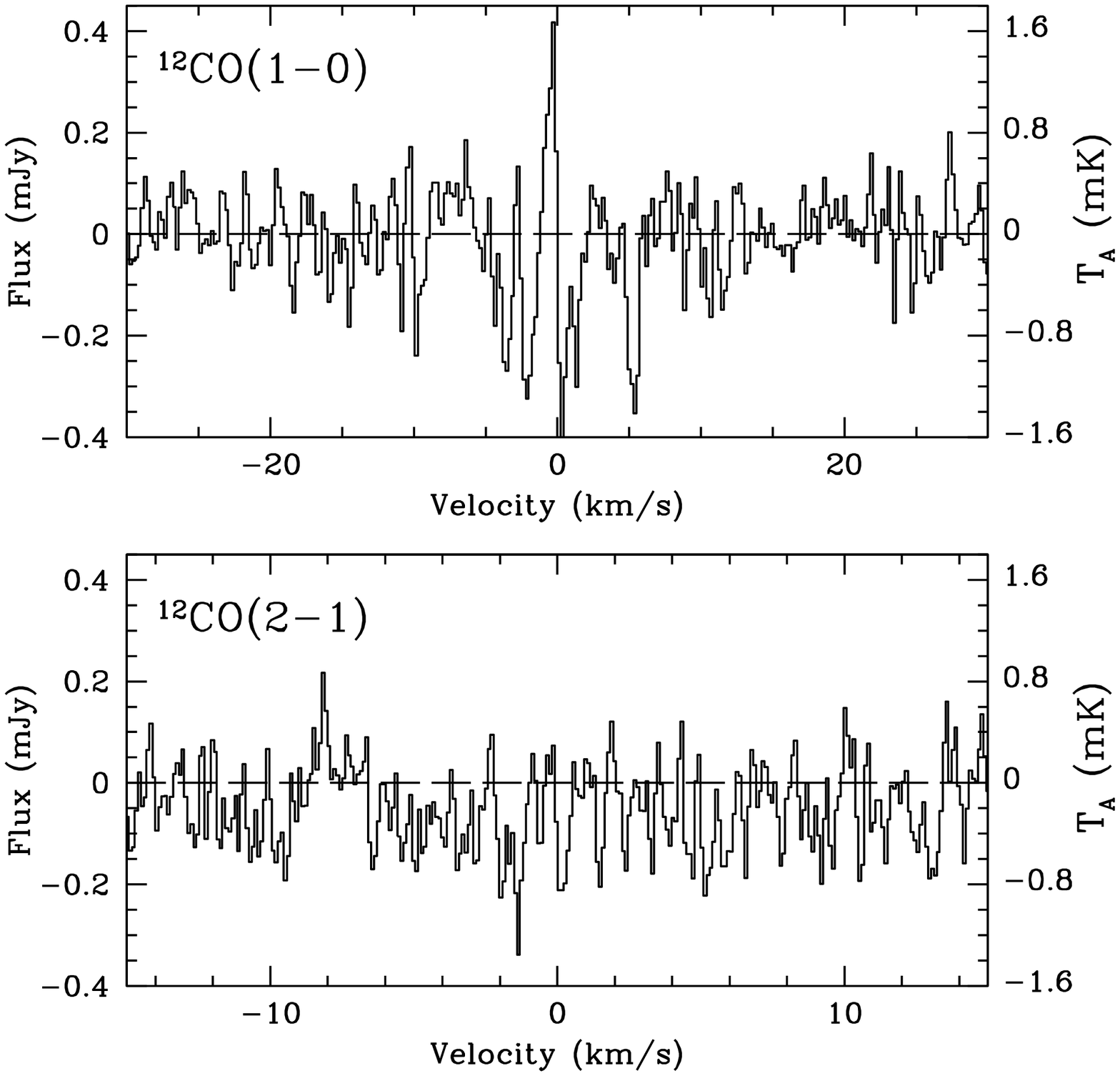}{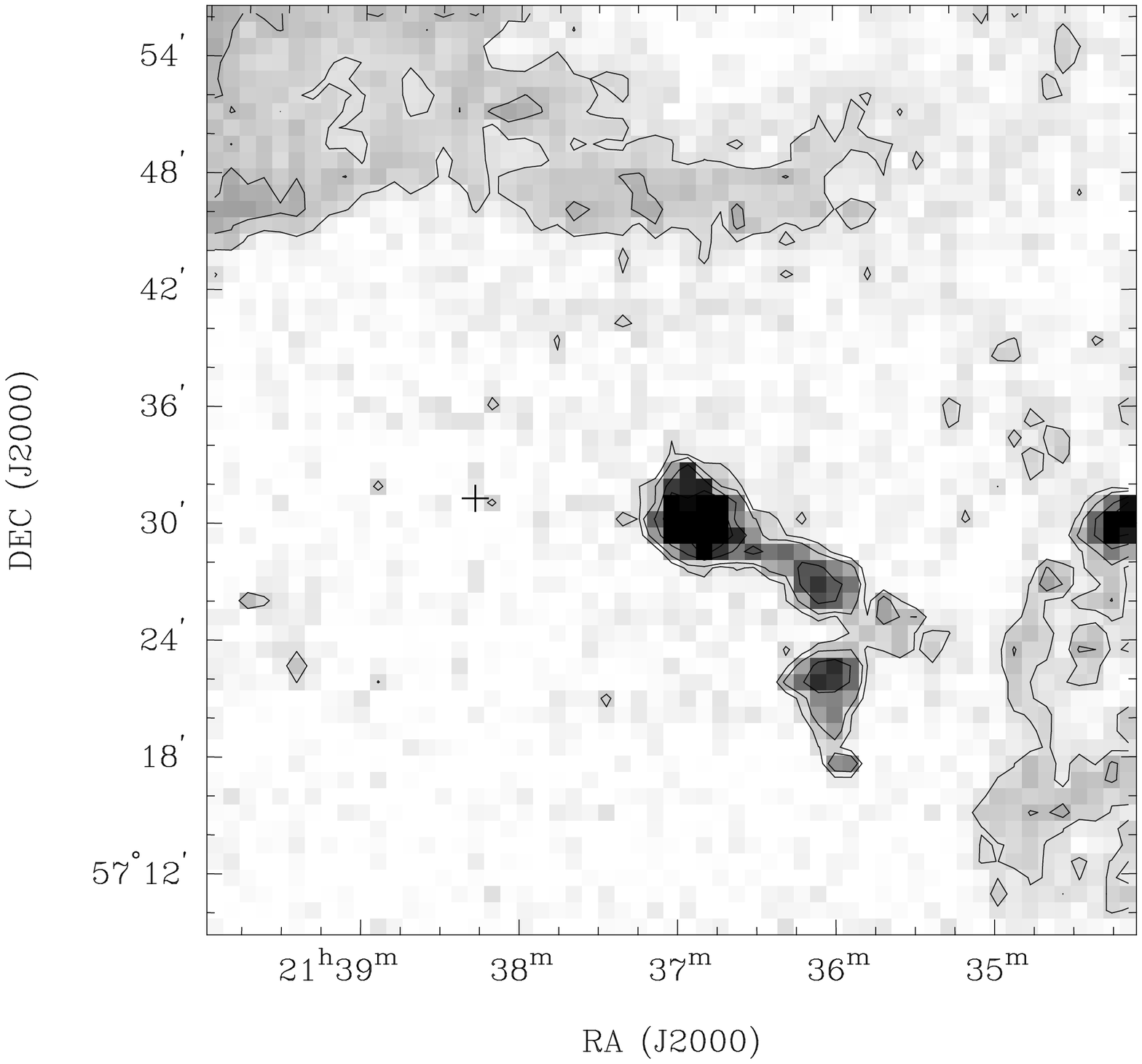}
\caption{Left: $^{12}$CO(1-0) and $^{12}$CO(2-1) spectra. The zero frequencies are
115.2712 and 230.538 GHz, respectively. A narrow emission is
seen at approximately the cluster velocity only in$^{12}$CO(1-0),
suggeting that the radial velocity of GM Cep is consistent with the radial 
velocity of the cluster. Right: FCRAO CO J=1-0 peak intensity map of the IC1396 region 
near GM Cep (Patel et al. 1998). The position of GM Cep is marked as a ``+''. The  
halftone greyscale levels are linearly scaled from 0.5 to 8 K, with  
contour levels: -3, 3, 5, 10, 15, 20... times the rms of the map (0.5  
K). \label{lines-fig}}
\end{figure}

The data were reduced using the CLASS software within the GILDAS package, 
specially developed for the reduction of IRAM heterodyne data. The single 
scans were examined to remove the bad ones, matched in frequency, and added. 
The baseline was fitted as a 1-order polynomial and subtracted. The reduced 
data is plotted in Figure \ref{lines-fig}. The $^{12}$CO(1-0) is identified
as a 5$\sigma$ detection with peak value 0.3mJy, but the weaker $^{12}$CO(2-1) 
transition is not detected. No line broadening was observed for the 
$^{12}$CO(1-0), which may be an effect of the weakness of the detection, 
and/or a consequence of the inclination of the object. In addition to the 
5$\sigma$ emission, we find a 4$\sigma$ absorption around the line. This 
absorption does not seem to originate in a single point in the off position, 
given the changes in the wobbler throw and its position angles. A possible 
explanation would be the presence of substantial amounts of gas in the Tr 37 bubble. 
According to the work of Patel et al. (1998) using the 14m FCRAO antenna, the region 
around GM Cep is relatively clean (Figure \ref{lines-fig} right). Nevertheless, the
sensitivity of the 30m is more than double than the FCRAO, so the absorption could 
be due to gas emission in the expanding IC 1396 H II region barely detected in the 
FCRAO maps. The $^{12}$CO lines are weaker than expected in typical disks for the
same 1.3 mm continuum emission, which could be due to oversubtraction of ambient 
$^{12}$CO emission, and is difficult to correct without deep mapping of the area.

\section{Analysis \label{analysis}}
 
\subsection{The Lightcurve: Outburst History \label{lightcurve}}

\begin{figure}
\plotone{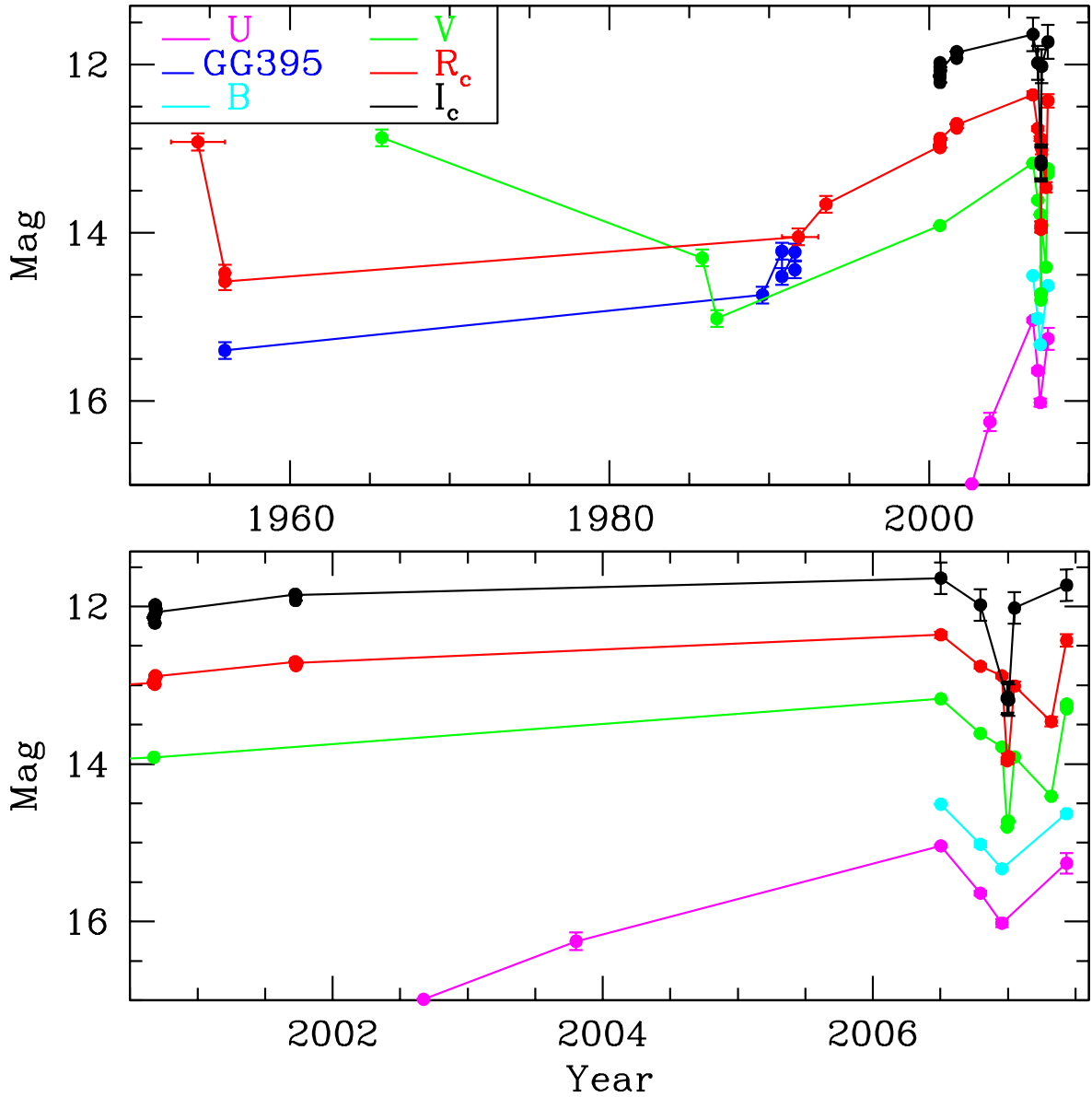}
\caption{Lightcurve for GM Cep, including the data in Table \ref{opt-table}.
The left pannel shows the complete lightcurve.
The right pannel shows a zoom on the most recent data.
The datapoints have been connected for better visualization, although due to the undersampling,
further magnitude oscillations probably occurred in between the measurements. \label{lightcurve-fig}}
\end{figure}

The lightcurve of GM Cep is depicted in Figure \ref{lightcurve-fig}. The star 
suffers rapid and repeated irregular variations in all bands, with amplitudes 
2-2.5 mags. It increased in brightness after $\sim$1986, reached a low magnitude 
level in $\sim$2002, and raised between $\sim$2003 and $\sim$July 2006. From 
July 2006 on, when our time sampling is better, GM Cep experienced a decrease 
in magnitude, followed by a 1 mag increase in January 2007, a further decline by 
April 2007, increasing again to near-maximum levels by June 2007. 
The changes in magnitude can be very rapid ($>$0.1 mag in consecutive days in 
2000 and 2001; 2 mag from Dec. 15 to  Dec. 29, 2006; 1 mag from 
Jan. 2 to Jan. 18, 2007). This suggests that GM Cep 
probably suffered similar rapid oscillations during the undersampled periods.

\begin{figure}
\plotone{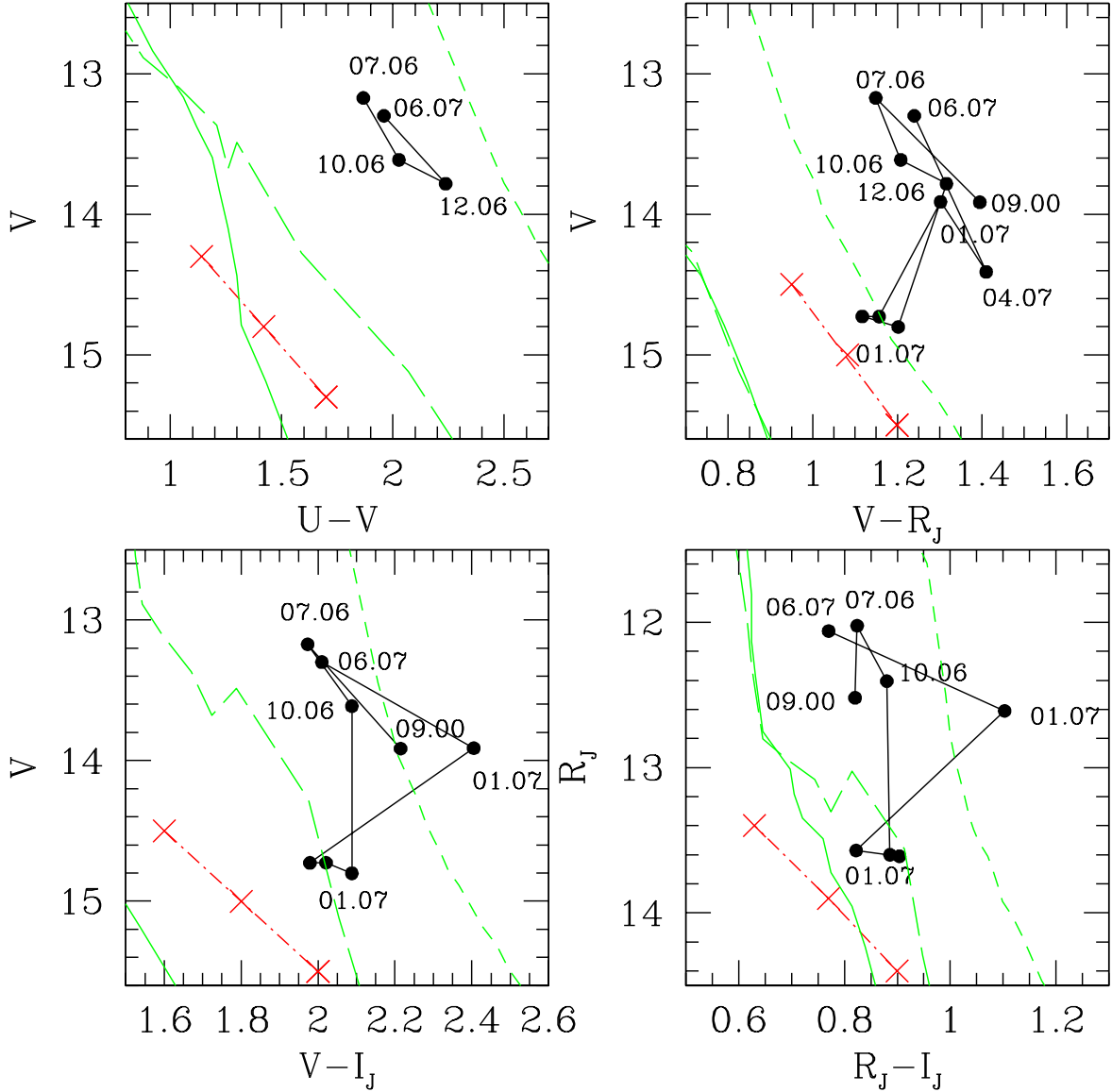}
\caption{Color evolution of GM Cep, compared to the 1, 10, and 100 Myr isochrones
(from right to left, short-dashed, long-dashed and thin lines; 
Siess et al. 2000) and to the reddening vector (dashed-dotted line marked at intervals
of A$_V$=0.5 mag). The data from September 2000
has been converted from the Cousins into the Johnson system (Fernie 1983).
The individual data points have been connected to show the direction of the variations.
Typical errors are smaller than the dots except for the September 2000 converted data,
which has errors up to $\sim$0.07 and $\sim$0.2 mags in R$_J$ and in I$_J$, respectively.
The isochrones have been reddened by the cluster average (A$_V$=1.67 mag) following
a standard reddening law (Cardelli et al. 1989).
 \label{colorJ-fig}}
\end{figure}

The simultaneous multiband data show that the color changes 
are small and relatively random (Figure \ref{colorJ-fig}). The changes in color do 
not seem  purely related to extinction variations, considering a typical interstellar 
extinction law (Cardelli et al. 1989), although the material surrounding a young star is 
likely to be different from typical ISM matter. If the changes in magnitude were due to 
extinction alone, we would need to assume significant grain growth and a non-standard 
extinction law: To produce the same amplitude of variations in all VRI bands, the obscuring 
grains should be very large ($>$10$\mu$m) to ensure grey extinction (Eiroa et al. 2002). 
Simultaneous JHK and optical observations would help to constraint the
effects of extinction in the lightcurve, but unfortunately only one of our JHK data
sets has simultaneous UBVRI photometry. The small color variations suggest as well that 
eclipses by a companion of different mass are not responsible for the changes. The amplitude 
of the variations (up to 2.5 magnitudes) is too large for eclipses, even if we consider 
an equal-mass companion, unless we consider some special binary configuration in addition 
to obscuring material, somehow analogous to KH 15D (Hamilton et al. 2005 and references therein). 
Detailed studies of the periodicity  in the future should clarify the presence of eclipses. 
Nevertheless, both the extinction and the eclipse scenarios fail to explain the high luminosity
of GM Cep (up to 30-40 L$_\odot$ at maximum) unless the star had a much earlier
spectral type (A) or belonged to another luminosity class.

\subsection{Stellar Properties and Activity \label{spectra}}

\begin{figure}
\plotone{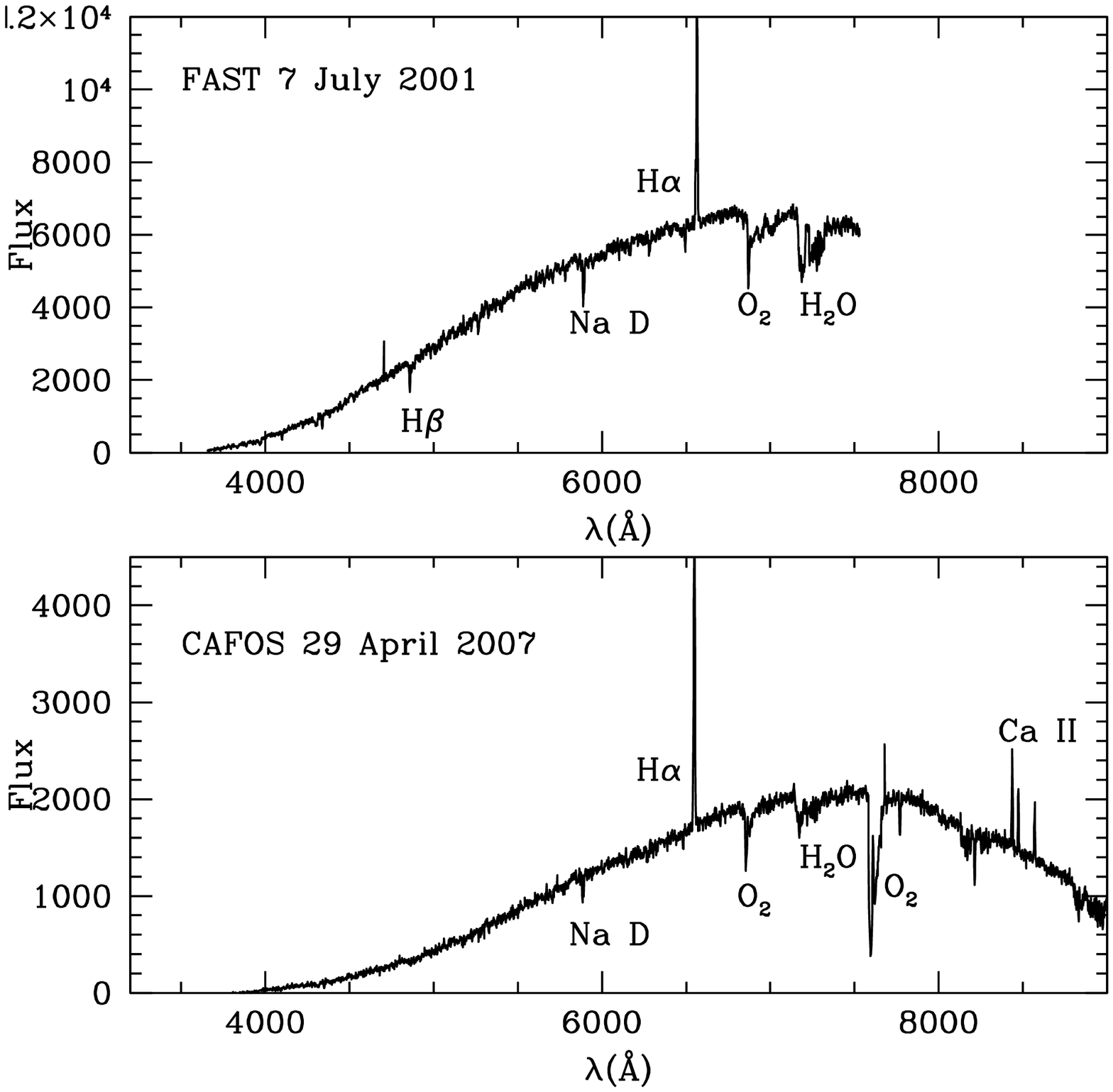}
\caption{Low-resolution spectra of GM Cep. The main emission and absorption
features, as well as the atmospheric H$_2$O and O$_2$ bands, are labeled.\label{lowrespec-fig}}
\end{figure}

\begin{figure}
\plotone{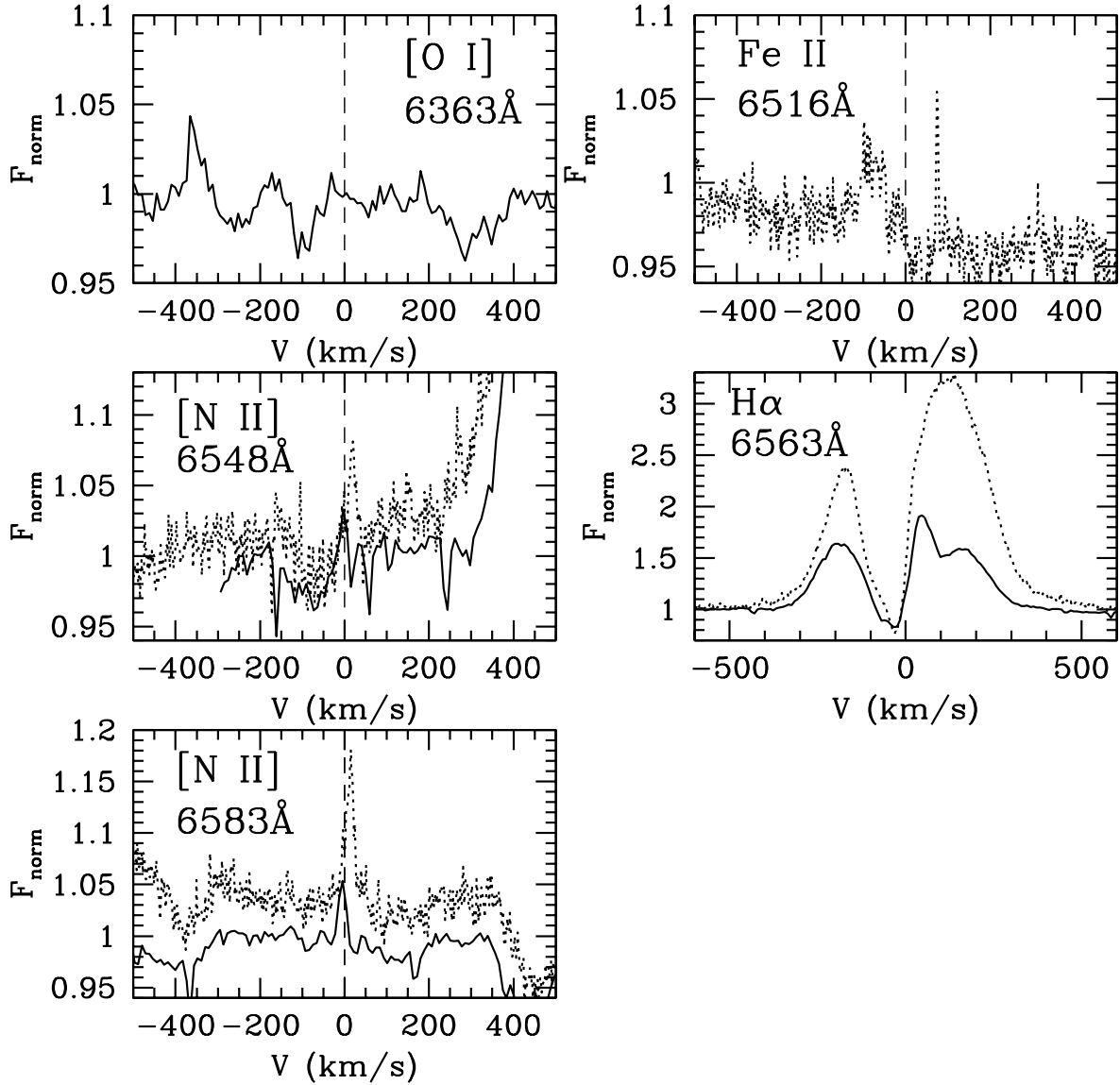}
\caption{Emission lines in GM Cep, from HIRES (solid line)
and from Hectochelle (dotted line) spectra. The zero velocity position has been marked
according to the HIRES radial velocity CZ=-21.0$\pm$2.6 km/s. \label{lines1-fig}}
\end{figure}

\begin{figure}
\plotone{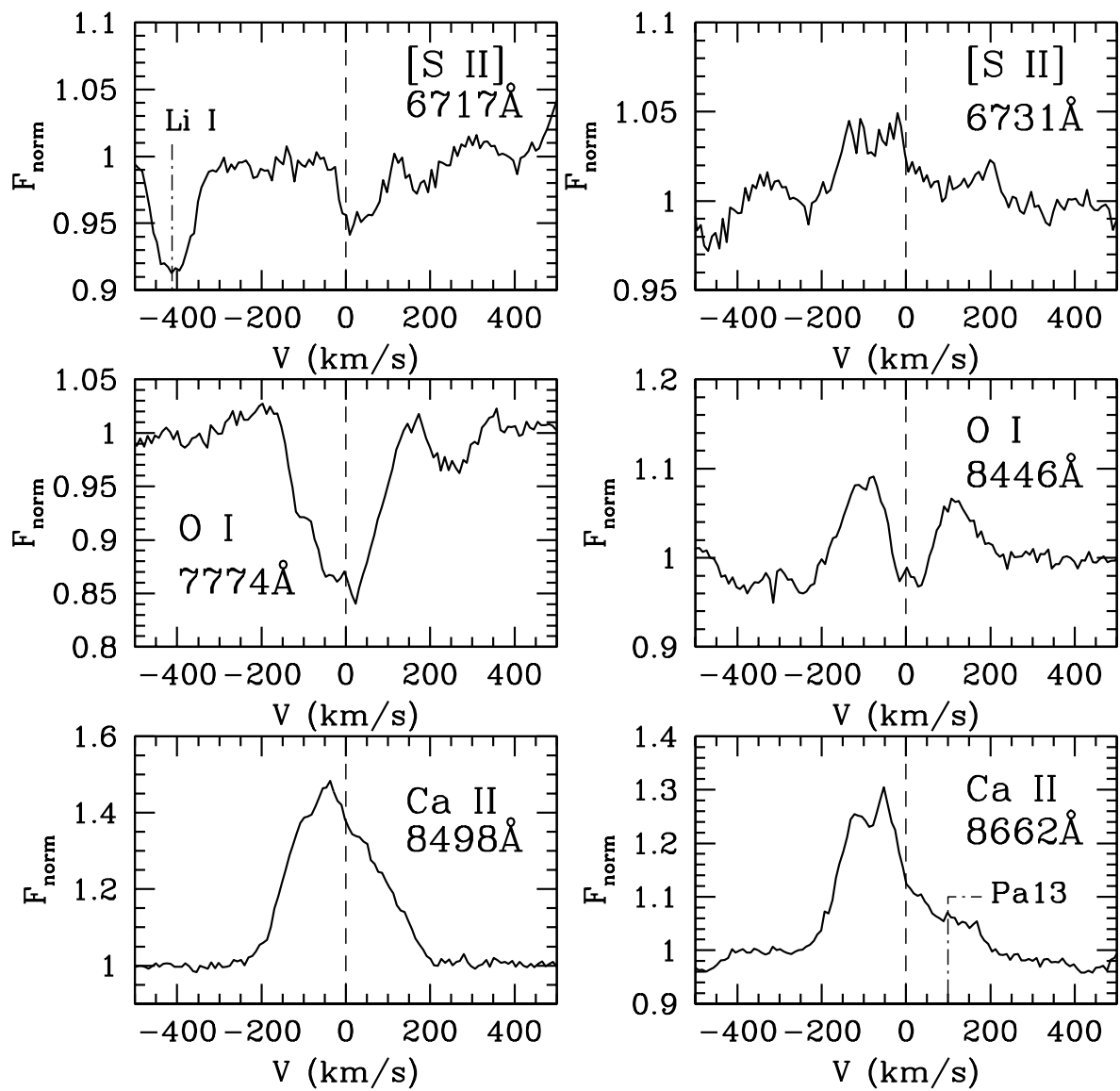}
\caption{Emission lines in GM Cep observed with HIRES. The zero velocity position has been marked
according to the HIRES radial velocity CZ=-21.0$\pm$2.6 km/s. The Li I absorption at 6708\AA\ and 
the Paschen 13 H line at 8663\AA\ have been marked by dotted-dashed lines.  \label{lines2-fig}}
\end{figure}

The spectra of GM Cep are dominated by very strong and broad H$\alpha$ emission 
with strong P-Cygni profile with a deep blueshifted absorption that goes under the 
continuum level, characteristic of very strong accretion (Edwards et al. 1994; 
Hartigan et al. 1995). Other emission lines associated to winds and accretion 
are present as well (Figures \ref{lowrespec-fig}, \ref{lines1-fig}, and \ref{lines2-fig}). 
The main lines are listed in Table \ref{lines-table}. Given that none of the spectra 
is flux calibrated, it is not possible to determine the variations in the continuum level. 
The spectral features in the high-resolution data from 2001 and 2004 are very similar. 
The H$\alpha$ equivalent width (EW) was larger in the 2004 (EW=-14 \AA) than in 2001 
(EW=-6 \AA), and the 10\% H$\alpha$ velocity width was also higher in 2004 (660 km/s; 
Paper IV) than in 2001 (580 km/s), suggesting a higher accretion rate in the first case. 
The narrow H$\alpha$ feature in the HIRES spectrum may be affected by nebular emission 
of the H II region, or may have been oversubtracted in the Hectochelle spectrum. The 
low-resolution spectra give H$\alpha$ EW -10 and -19 \AA\ in June 2001 and April 2007, 
respectively, suggesting H$\alpha$ variability caused by variable accretion/winds added to
changes in the continuum level. 

In the 2001 low-resolution FAST spectrum, the H$\beta$ line appears in absorption.
Typically, accreting stars show the Balmer H Series in emission, with similar 
profiles in all the lines, suggesting that they came from the same mass of emitting gas
(Muzerolle et al. 1998, 2001). Nevertheless, in the case of very strong
accretors, the strong winds can dominate the most optically thick lines
(H$\alpha$) overrunning the signs of magnetospheric accretion, which would
dominate optically thinner lines (like H$\beta$; Muzerolle et al. 2001). 
Intermediate-mass variables of the UX-or class tend to have H$\beta$ in absorption
(Grinin et al. 2001; Tambovtseva et al. 2001), but H$\alpha$ P-Cygni profiles
are very rare in them, and the Ca II IR lines appear typically in absorption.
For GM Cep, the Ca II IR triplet ($\lambda \lambda$ 8498, 8542, 8662 \AA), indicator 
of accretion, shows strong emission in both the HIRES and the CAFOS spectra 
(Figures \ref{lowrespec-fig} and \ref{lines1-fig}). The lines were stronger in 2001, 
and the ratio of the 8498 and the 8662 line varies in the two epochs from $\sim$1.5 
to $\sim$2.5. The profiles of these two lines are not similar either, but the 8662 \AA\ 
line can be blended the Paschen 13 line. 

The high-resolution spectra show forbidden [N II] emission at 6548 and 6583 \AA, 
typically associated to winds and shocks (Hartmann \& Raymond 1989). The [N II] lines 
are stronger in the 2004 data, which has the stronger H$\alpha$. In addition, the [N II] 
lines appear redshifted by about 18 km/s in 2004 with respect to 2001. The shock origin 
of the [N II] emission results typically in small blueshifts of the lines 
(Hartigan et al. 1995), so the velocity shift observed could be related to changes 
in the velocity and/or location of the shocked matter, or in the radial velocity
of the star. Considering the strength of the lines, we believe that GM Cep had stronger 
winds during 2004 than in 2001. The [O I] line at 6300 \AA\ is  not covered, and the [O I] 
line at 6363\AA\ is not detected, but the [S II] line at 6729 \AA\ is detected in emission.
The forbidden emission lines ([O I], [N II], [S II]) can be used to constrain the wind 
and shock density and the temperature of the regions where they originate, given their 
different critical densities and ionization potentials (Hartigan et al. 1990). If a jet 
or wind does not encounter matter in its way, the forbidden line emission may not be 
produced (Hartigan et al. 1995). Higher densities tend to reduce the [N II] and [S II] 
emission compared to [O I], and higher shock velocities increase ionization and the 
[N II]/[O I] ratio. The different velocities observed for the different lines suggest 
different formation volumes: [N II] could be produced in a low-density shock far from 
the star, whereas the [O I] line could be related to denser, hotter wind maybe
associated to the material producing the H$\alpha$ absorption.

GM Cep presents a remarkable, double-peaked O I emission at 8446 \AA, with peaks 
centered at 8442 and 8448 \AA, respectively ($\sim$-140 km/s and $\sim$+80 km/s; 
Figure \ref{lines1-fig}). This line is characteristic of very strong winds, 
being usually present in Herbig Ae/Be stars and in strongly accreting TTS like RW Aur 
(Alencar et al. 2005). The double peak observed in GM Cep could in fact indicate two 
jet-like wind components. There is a prominent and broad O I absorption at 7774\AA, 
maybe related to the hot disk (Grinin et al. 2001). We detect as well some weak 
Fe II and Fe I lines in emission  (Table \ref{lines-table}), which are seen in 
spectra of other strongly accreting stars like V1647 (Fedele et al. 2007).

The new data allow to revise the spectral type of the star. The low-resolution spectra are  
veiled and lack broad band features, giving an uncertain spectral type around G. The 
high-resolution HIRES spectrum shows numerous photospheric lines in the whole wavelength 
range, despite important blending due to the fast rotation of GM Cep. Comparing to standard 
spectral libraries (Montes et al. 1997; Coluzzi et al. 1993), the Fe I and Ca I lines, 
together with the Li I absorption (0.16\AA) suggest a late-type in the range G7V-K0V 
(1$\sigma$; G5-K3 for 3$\sigma$). The main contributors to the uncertainty in the spectral 
type are the moderate veiling ($\sim$0.2-0.4, depending on spectral type) and the fast 
rotation, added to the intrinsic lack of strong features in G-type stars. 

\begin{figure}
\plotone{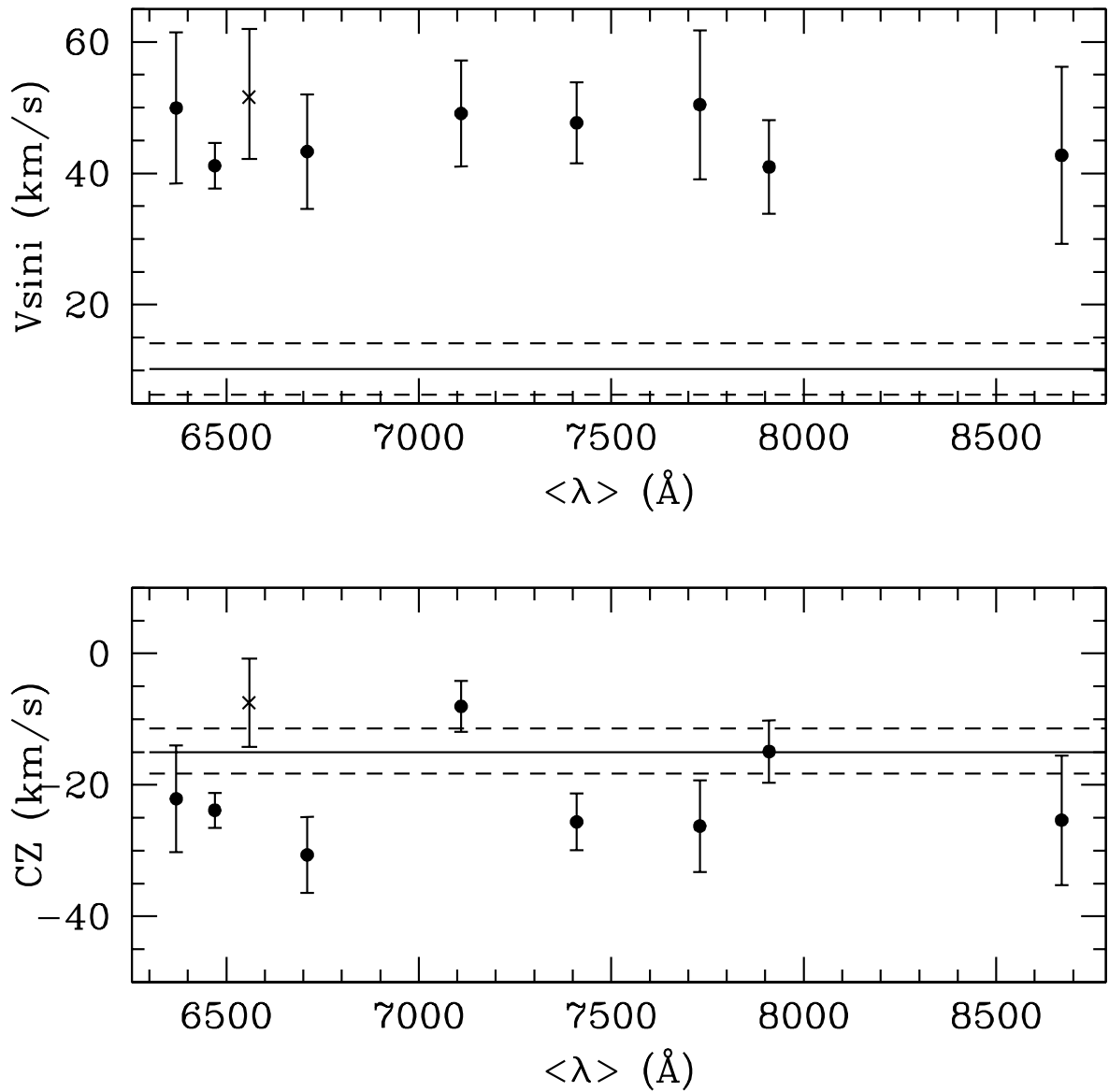}
\caption{Rotational velocity (V$sini$) and radial velocity (CZ) of GM Cep versus
wavelength, derived from the HIRES spectra (dots) and from Hectochelle data (crosses).
The average V$sini$ and CZ of the Tr 37 members are displayed as a solid line,
together with the 1$\sigma$ deviations seen in the cluster (dashed lines).
Whereas the radial velocity of GM Cep is consistent with the Tr 37 members within
the errors, the V$sini$ is much higher than any typical CTTS or WTTS. \label{xcor-fig}}
\end{figure}

Cross-correlation of the Hectochelle spectrum with spectra of slow rotators of a similar 
spectral type revealed a rotational velocity V$sini$=51.6$\pm$9.4 km/s (Paper IV). We repeated 
the cross-correlation with each one of the HIRES orders, removing the zones affected by 
emission lines, bad pixels, and strong atmospheric features (mostly O$_2$ and H$_2$O bands; 
Curcio et al. 1994). The results of the cross-correlation are summarized in Table \ref{xcor-table}, 
and displayed in Figure \ref{xcor-fig}. The HIRES data confirm the very high rotational velocity
over a large range of wavelengths, V$sini$=43.2 $\pm$ 1.6 km/s in average (standard deviation 
$\sigma$=4.2 km/s). Compared to the average rotational velocity of the rest of Tr 37 members 
(V$sini$=10.2$\pm$3.9 km/s), GM Cep is a very fast rotator: the next fast 
rotator in Tr 37 is the M0 weak-lined TTS 21-763, with V$sini$=28.3$\pm$8.1 km/s. We do not 
observe any correlation between V$sini$ and the wavelength (Figure \ref{xcor-fig}) nor between 
the line broadening and the line transition lower excitation potential, as has been 
suggested for FU-ors (Hartmann \& Kenyon 1987; Welty et al. 1990, 1992; Kenyon \& Hartmann 1996). 
Therefore, we believe that the absorption lines arise rather from the stellar photosphere than 
from a hot disk or shell. The cross-correlation provides as well a measure of the radial 
velocity. The average radial velocity in Tr 37 is CZ=-15.0$\pm$3.6 km/s, so GM Cep is slightly
off-cluster with CZ=-7.5$\pm$6.7 km/s (Paper IV). The HIRES data produces a different off-cluster
value, CZ=-21.0$\pm$2.6 km/s. On the other hand, the radial velocity of the $^{12}$CO (1-0) 
peak is consistent with the cluster average -15$\pm$2 km/s (Section \ref{data-co}).
Nevertheless, the broad lines lead to large errors, so the evidence of spectroscopic binarity 
is only 2$\sigma$ and therefore, not conclusive.

\subsection{Accretion \label{accretion}}

The U band observations combined with simultaneous VRI photometry can be used to 
determine the accretion rate via the excess U band luminosity (L$_U$; Gullbring et al. 1998; 
Paper I, II, IV). The excess in U band luminosity is well correlated to the accretion
luminosity L$_{acc}$, which can be transformed into an accretion rate. Following
Gullbring et al. (1998):

\begin{equation}
\log(L_{acc}/L_\odot) = 1.09 \log(L_U / L_\odot) + 0.98
\end{equation}

\begin{equation}
L_{acc} \sim G M_* \dot{M} / R_* (1-R_* / R_{in}) 
\end{equation}

The mass and radius of the star (M$_*$, R$_*$) can be obtained from the V vs. V-I
diagram and evolutionary tracks (Siess et al. 2000). L$_U$ can be estimated as 
the difference between the measured U band luminosity and the
photospheric luminosity, calculated from the measured I magnitude and using the U-I 
color of a non-accreting star with the same spectral type (Kenyon \& Hartmann 1995). 
All magnitudes must be corrected from extinction (derived from the VRI colors using a
standard galactic extinction law; Cardelli et al. 1989), and the transformation of 
magnitudes to luminosity can be done with the zero point flux and the bandwith 
for U band (4.19 10$^{-9}$ erg s$^{-1}$ cm$^{-2}$ \AA$^{-1}$ and 680 \AA, respectively). For 
GM Cep, and for an spectral type G7-K0, the extinction is A$_V$$\sim$2-3 mag, so the 2006-2007
data results in variable L$_U \sim$0.5 - 4 L$_\odot$ and a variable accretion rate in 
the range \.{M}$\sim$10$^{-7}$ - 5 10$^{-6}$ M$_\odot$/yr. The mass and radius of the star 
are uncertain, given the variations within the color-magnitude diagram (Figure \ref{colorJ-fig}), 
being the most probable values M$_*$=2.1 M$_\odot$ and a radius between
3 to 6 R$_\odot$. Added to the uncertainty in the spectral type and
extinction, this result is accurate within 1 order of magnitude. In any case, 
this accretion rate is 2-3 orders of magnitude higher than the median in Tr 37 
(Paper IV), and 1-2 orders of magnitude higher than the rates in typical Taurus stars.  

Using the parametrization in Natta et al. (2004), we can convert the H$\alpha$ velocity 
wings at 10\% of the maximum (V$_{H\alpha 10\%}$) into accretion rates:

\begin{equation}
\log(\dot{M} / M_\odot yr^{-1}) = -12.89 + 9.7 \times 10^{-3} V_{H\alpha 10\%} / km s^{-1}
\end{equation}

For the measured velocities of 580 and 660 km/s, the accretion rates are 5 10$^{-8}$ M$_\odot$/yr 
and 3 10$^{-7}$ M$_\odot$/yr, respectively. These values may be underestimated given the wind
absorption in the H$\alpha$ profiles and depending on the viewing angle. 
As the dispersion associated to this parametrization is around 1-2 orders of magnitude, the 
3 10$^{-7}$ M$_\odot$/yr rate derived from Hectochelle in 2004 is roughly consistent with the 
values derived from U band photometry. The moderate veiling estimated from the HIRES spectrum
 ($\sim$0.2-0.4) agrees with the accretion rate estimated from H$\alpha$ assuming an intermediate 
luminosity for GM Cep.

\subsection{IR Variability and Disk Models \label{diskmodels}}

In order to study the structure of the disk around GM Cep, we trace the spectral energy 
distribution (SED) using the millimeter, Spitzer, 2MASS, and optical data. The main 
uncertainty here is that the different observations are not simultaneous. 
The 2MASS JHK data are nearly contemporary to the optical photometry (2000), but the Spitzer 
data were taken in 2003-2004, and the millimeter continuum in 2006. Short time variations of 
$\sim$1 mag in JHK have been observed for stars similar to GM Cep (Eiroa et al. 2002), 
suggesting structural changes in the disks. According to the  D'Alessio et al. (2005) disk 
models, the flux at wavelengths shorter than $\sim$ 300 $\mu$m scales linearly with the total 
(stellar and accretion) luminosity, but the disk flux at longer wavelengths is independent 
(Mer\'{\i}n et al. 2004). Juh\'{a}sz et al. (2007) and K\'{o}sp\'{a}l et al. (2007) found 
similar evidences of mid-IR excess variations with luminosity. If the causes of the outbursts 
were extinction variations, the mid- and far-IR variability would be milder, and the near-IR 
would be only moderately variable.  The differences between the fluxes measured by IRAS 12$\mu$m, 
IRAC 8$\mu$m, and the MSX6C A band ($\sim$8 $\mu$m) are suggestive of IR variability. On the 
other hand, the millimeter emission reflects the total mass of the disk, changing little even 
if the star suffers strong variability.The brightness in V varied in the whole range 
($\sim$2-2.5 mags) during the time when the different IR observations were taken 
(Figure \ref{lightcurve-fig}), so for modelling the disk we take the optical data from
September 2000, which has an intermediate brigthness, and keep in mind that the SED
may be affected by variable IR fluxes.

\begin{figure}
\plotone{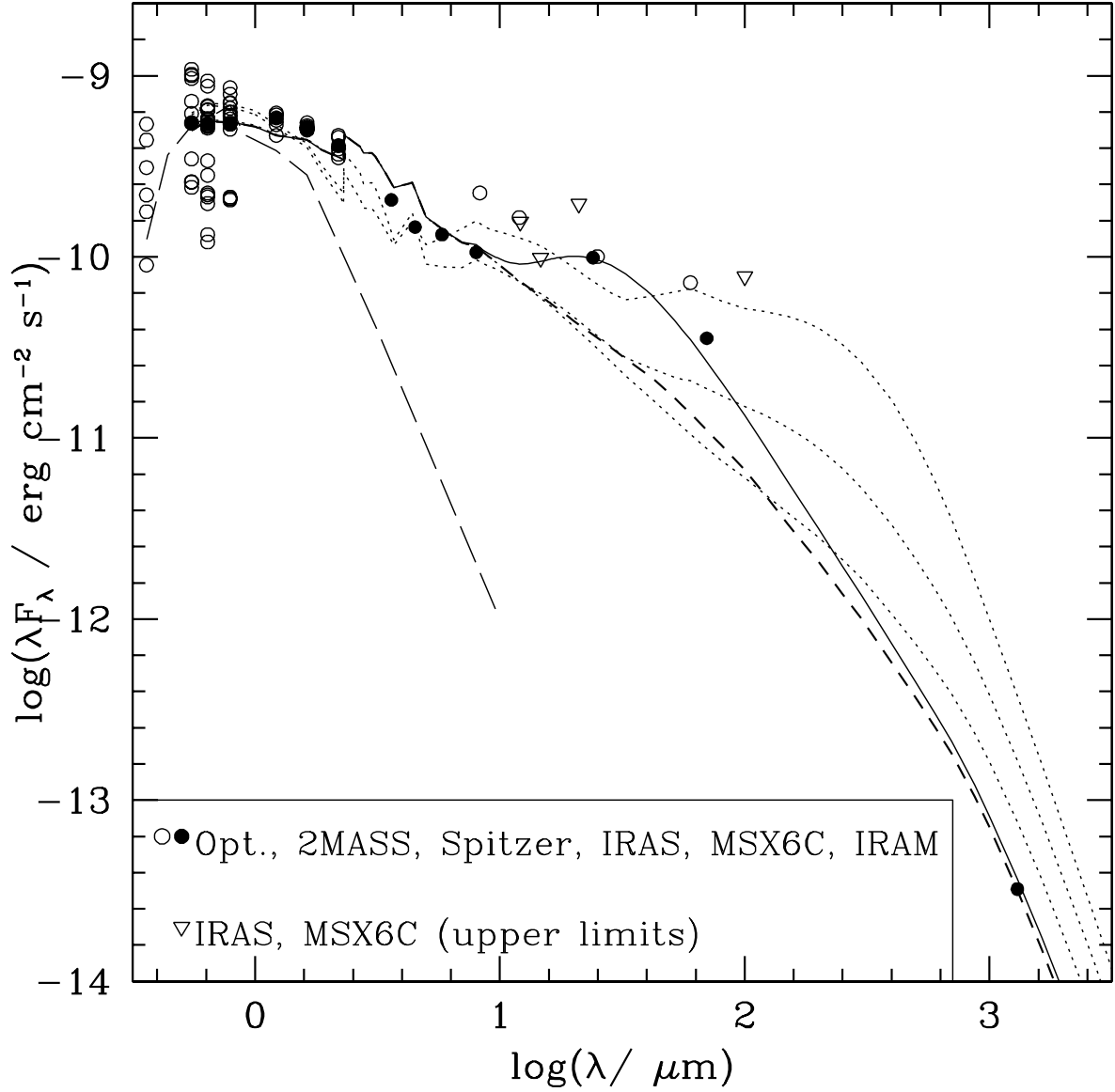}
\caption{SED data and disk models (D'Alessio et al. 2005) for GM Cep. The fitted data 
points (see text) are represented by filled circles. Open circles mark the rest of data 
points not included in the fitting, and inverted triangles are upper limits. The dotted 
lines trace three different hydrostatic equilibrium disk models for a star with luminosity 
26 L$_\odot$, accretion rates 10$^{-7}$-10$^{-6}$ M$_\odot$/yr, and disk radii ranging from 
50 to 600 AU. The short dashed line represent the model that best fits the millimetre point,
consisting of a disk with \.{M}$\sim$10$^{-6}$ M$_\odot$/yr and 30 AU outer radius, and the
bold line is the 30 AU model plus a black body with T=150 K to simulate an excess flaring
or envelope in order to reproduce the high mid-IR fluxes. The long dashed line is the photospheric 
emission for a K0 star (Kenyon \& Hartmann 1995). The data is de-reddened for A$_V$=2.5 mag.
The unability to fit simultaneously the mid-IR (24 and 70$\mu$m) and the millimeter point
with the standard equilibrium models suggests that the disk is unstable and has probably
suffered considerable grain growth. \label{bestfit-fig}}
\end{figure}

Typical disk models assume that the disk is in hydrostatic equilibrium. Nevertheless, the 
strong IR emission suggests that the disk is compact, dense, and, if the variability is triggered 
by episodes of enhanced accretion, probably out of hydrostatic equilibrium (Clarke \& Lin 1990; 
Eiroa et al. 2002), so we must be cautious interpreting the models. We use the standard 
models for irradiated accretion disks around TTS of D'Alessio et al. (2005), which trace the 
flared disk self-consistently, with a flaring angle determined by the hydrostatic equilibrium, 
and include stellar illumination and accretion as heating sources for the disk. The dust 
opacities are calculated for a collisional distribution of 
grain sizes, where the amount of grains of a given size ``a" varies as a power law
n(a)$\propto$ a$^{-3.5}$ and the maximum grain size is 10$\mu$m. We assume an average inclination 
of 60 degrees; the effects of intermediate inclination variations being only important for the 
near-IR emission. The stellar parameters, derived from the optical photometry of September 2000
using the HR diagram, are M=3.0 M$_\odot$, R=5.14 R$_\odot$, T=5000 K. We also include an accretion 
rate of  10$^{-6}$ M$_\odot$/yr. These parameters are one of the main uncertainties in the modeling, 
given the variability of the star. The best fit model is displayed in Figure \ref{bestfit-fig}.  
The result is, not surprising, that the disk cannot be modeled as a standard disk. If we try to fit 
the MIPS data points at 24 and 70 $\mu$m, the 1.3mm continuum flux would be overestimated by more 
than one order of magnitude, and even the 24 and 70 $\mu$m contemporary data alone cannot be 
accurately reproduced with hydrostatic equilibrium disk models for any observed stellar luminosity. 
The mid-IR flux of GM Cep is comparable to that of FU Orionis itself (Quanz et al. 2006), 
although its disk (or disk plus envelope) is less massive. Since the flux at millimeter wavelengths 
is independent of the rapid optical variability, a standard hydrostatic equilibrium model can 
reproduce the flux at 1.3mm only if the grains are either much smaller or much larger than 
10 $\mu$m (maximum grain size $<$1$\mu$m or $>$1cm) or if the disk is very small (up to 30-50 AU). 
Even in these cases, an additional source of IR emission should be included to produce the high 
24 and 70 $\mu$m flux (in a small disk) or to fit the near-IR excess (if the grain sizes have 
cm sizes). 

A way to increase the IR fluxes without changing the mm flux would be by  varying the 
flaring angle at a certain distance from the star. The flaring angle in the 
D'Alessio et al. (2005) models is fixed by the hydrostatic equilibrium, but
strong turbulence related to the strong accretion can produce an unstable
disk. As an experiment, the mid-IR fluxes can be increased by adding a black body 
emission with T$\sim$150 K to a compact 30 AU disk model. For the luminosity of 
GM Cep, this would mean a distance of $\sim$ 25 AU. The excess mid-IR emission that 
needs to be added could be interpreted as an increase in the flaring angle by a factor
of $\sim$2 at distances of $\sim$20-30 AU (a ``bump'' in the disk), as an illuminated 
wall at $\sim$25 AU, or as a shell or envelope located at a similar distance. If the 
innermost disk were very flared or had a high inner wall, some degree of self-shadowing 
in the near-IR may occur, leading to a ``bump'' at longer wavelengths (Dullemond et al. 2001).

A physical mechanism to produce a ``bump'' or a local change in the flaring angle could 
be to disrupt the disk by a massive object (a stellar companion and/or a giant planet) 
triggering thermal instability and producing rims and/or spiral waves (Clarke \& Lin 1990; 
Clarke \& Syer 1996; Lodato \& Clarke 2004). This picture would also be useful to explain 
the observed outbursts as periods of increased accretion activity. A close-in companion, 
a companion embedded in the disk, or a companion out of the disk (typically, at distances
2-3 times the size of the disk; Artymowicz \& Lubow 1994) could produce disk instabilities 
and increased accretion episodes, with the advantage that companion-triggered outbursts can 
occur at later stages of evolution in Class II objects, more consistent with the ages of Tr 37
(Lodato \& Clarke 2004). Detailed studies of the periodicity and radial velocity will be 
used to test these scenarios.

The shell/envelope picture presents a main problem: the relatively low extinction of GM Cep. 
Assuming the reasonable spectral type range (G5-K3) the extinction can vary between A$_V$=1.5-3 mag, 
roughly consistent with the cluster average A$_V$=1.67$\pm$0.45 mag (Paper II), and with the 
minimum interstellar extinction for an object located at 900 pc, A$_V \sim$1 mag.
Therefore, the column density in the envelope would be of the order of N(H$_2$)=10$^{21}$ cm$^{-2}$ 
at most (Frerking et al. 1982; Taylor et al. 1993), resulting in a total mass of 
$\sim$4$\times 10^{-5}$M$_\odot$ if the envelope has a $\sim$30 AU radius, and still 
only 0.001 M$_\odot$ for a $\sim$500 AU radius, insignificant compared to the total
disk mass. The envelope could be more dense and massive if it had polar gaps, which is 
reasonable given the strong winds, and if we assume a low viewing angle, but it cannot be 
too massive to reproduce the millimeter flux. A low viewing angle would be consistent with 
the lack of broadening in the $^{12}$CO(1-0) line, and would suggest that GM Cep is rotating close
to the breakup velocity, $\sim$200 km/s. Accretion from an envelope onto the disk could produce 
episodes of increased accretion onto the star. Extinction by a non-uniform envelope 
can produce magnitude changes as well, but the rapid variation timescales require that, if 
extinction plays an important role in the magnitude changes, the extincting matter must be 
very close to the star. In order to distinguish between all these scenarios, simultaneous 
observations at all wavelengths are required.

\subsection{Disk Mass Estimates \label{diskmass}}

The millimeter continuum data can be used to estimate the
total disk mass, given that the whole disk is optically thin
at these wavelengths. The total dust mass is written as:

\begin{equation}
M_{dust} =  F_\nu D^2 / \kappa_\nu B_\nu(T_{dust})
\end{equation}

where F$_\nu$ is the flux at millimeter wavelengths, D is the distance, 
$\kappa_\nu$ is the dust mass absorption coefficient at the same wavelength
(2 cm$^{2}$ g$^{-1}$ at 1.3mm; Beckwith et al. 1990; Miyake \& Nakagawa 1993; 
Kr\"{u}gel \& Siebenmorgen 1994). B$_\nu$(T$_{dust}$) is the Planck function 
for the temperature of the bulk of the dust. The total mass of the disk can be 
obtained assuming a gas-to-dust ratio of 100 (Beckwith et al. 1990). The main 
uncertainties are the dust absorption coefficient (if substantial grain growth
has occurred), the gas-to-dust ratio (if the disk is older than typical disks), 
and, in an unstable disk, the temperature of the bulk of the dust. We use the 
dust absorption coefficient assumed for Taurus disks (Beckwith et al. 1990; 
Osterloh \& Beckwith 1995), which is larger than the $\kappa_\nu$ for the interstellar
medium (1 cm$^{2}$ g$^{-1}$) or than the one of disks with grains larger than $\sim$100 $\mu$m 
(Kr\"{u}gel \& Siebenmorgen 1994). The typical dust temperature for outer disks around TTS is 
30-40 K (Dutrey et al. 1997; Thi et al. 2001). For GM Cep, the dust temperature could be 
higher if the strong accretion is adding extra heating to the disk, and/or if the bulk of 
the dust is closer to the star (because of changes in the flaring and/or envelope configuration). 
In order to be conservative, we consider dust temperatures in the range 30-150 K, 
resulting in total disks masses from 0.07 to 0.01 M$_\odot$, several times the 
minimum mass solar nebula.

The emission of optically thin transitions in molecules can be used to estimate the
mass in cold gas (Thi et al. 2001). The $^{12}$CO(1-0) line detected at the IRAM 30m 
telescope is most likely optically thick, so this would  underestimate the disk mass. 
Nevertheless, if we make the approximate calculation, assuming that the line is 
thermalized, we have:

\begin{equation}
F_{(1-0)} = (h \nu / 4 \pi D^2) N(CO) A_{1-0} \chi_\nu A
\end{equation}

where F$_{(1-0)}$ is the total flux in the $^{12}$CO(1-0) line,
N(CO) is the $^{12}$CO column density, D is the distance, A$_{1-0}$ is the
Einstein coefficient for the transition,  $\chi_\nu$ is the 
level occupation at a given temperature, and A is the disk surface area. 
The main error in this approach is that the line is not optically thin. 
If we assume the ratio of H$_2$ to $^{12}$CO to be 10$^4$
(Thi et al. 2001), and we consider the total flux integrated 
over the line profile F$_{(1-0)}$, the mass of the disk can be written as:

\begin{equation}
M_{disk}= \mu m_{H_2} F_{(1-0)} 4 \pi D^2 (N(H_2)/N(CO))/ h \nu A_{1-0} \chi_\nu
\end{equation}

With A$_{1-0}$ and $\chi_\nu$ from the Leiden Atomic and Molecular Database (LAMBDA; 
Sch\"{o}ier et al. 2005) and the Cologne Database of Molecular Spectroscopy (CDMS; 
M\"{u}ller et al. 2001, 2005), and assuming gas temperatures between 40 and 80 K 
(this variation results only in a factor of 2), the total disk mass inferred from the 
integrated flux F$_{(1-0)}$=0.27 Jy km/s is of the order of 0.0005 M$_\odot$,
about 100 times smaller than the estimate from the dust
continuum emission. This is reasonable since the line is optically thick, and 
the typical CO depletion values in TTS disks are of the order of 10-1000 
(Thi et al. 2001). Moreover, our $^{12}$CO fluxes may be underestimated
because of the absorption from ambient gas (Section \ref{data-co}).

The disk mass obtained integrating the 30 AU disk model that gives the best
fit to the millimeter point (Section \ref{diskmodels}) is about 3 times larger,
0.19 M$_\odot$. The model suggests a temperature around 30-40 K for the bulk
of dust. The main source of disagreement is the dust 
mass absorption coefficient, which in the model is variable with the radius of 
the disk and depends on the distribution of grains of different sizes, having
typical values between 0.4 and 1.8 cm$^2$ g$^{-1}$, smaller than the $\kappa _\nu$
from Beckwith et al. (1990). The stronger inconsistency is that for an accretion 
rate \.{M}$\sim$10$^{-6}$M$_\odot$/yr, the disk would survive less than 
0.1 Myr, very little compared with the ages of Tr 37 and even with the ages of the Tr 37 
globule. This inconsistency is not exclusive to GM Cep, but many of the strong accretors like 
DR Tau and RW Aur have similar \.{M} and even smaller millimeter disk masses
(0.03 and 0.0003 M$\odot$, respectively; Beckwith et al. 1990; Osterloh \& Beckwith 1995), so 
their survival times would be even shorter. In case of RW Aur, the problem can
be solved assuming that the current accretion episodes are transient events caused by 
disk disturbances by its companion(s) (Petrov et al. 1991; Cabrit 2005). EX-ors are 
supposed to have variable accretion, so they could spend most of their lives accreting at 
lower rates. Another way to solve the disagreement would be to change the opacities by 
increasing the maximum grain sizes to $\sim$1-10 cm or decreasing it under 1$\mu$m. Such dust
distributions are compatible with the millimeter fluxes if the disk has a mass 
$\sim$0.5-0.6 M$_\odot$ and outer radius $>$100 AU. Mannings \& Emerson (1994) suggested that 
including large grains and fractal dust structures could increase the disk mass of DR Tau, 
and Rodmann et al. (2006) found evidence of cm-sized grains in RW Aur.
Important grain growth seems appropriate for older TTS like GM Cep and should
be explored with more observations in the submm range to constrain the SED slope.

\subsection{Does GM Cep have a Wide-Orbit Companion? \label{companion}}

Given the anomalous SED of GM Cep, the repeated outbursts, and the fact that many of 
the well-known variable and EX-or/FU-or objects are multiple systems with separations
$\sim$100-300 AU (Reipurth \& Aspin 2004; Vittone \& Errico 2005; Ghez et al. 1993), we 
investigate the presence of companions. The presence of binaries has been invoked as a 
mechanism to produce EX-or and FU-or outbursts in old and more evolved disks without 
an envelope (Clarke \& Lin 1990; Bonnell \& Bastien 1992). Some very 
variable TTS like RW Aur are such multiple systems (Ghez et al. 1993), and companions at 
distances $\sim$100 AU affect the evolution of disks around Herbig AeBe stars 
(Chen et al. 2006) and TTS (Bouwman et al. 2006). The high-resolution spectroscopy is not 
conclusive, but the detection of spectroscopic binaries may be difficult if the star is 
a very fast rotator, and/or if the viewing angle is low. The lucky imaging data from 
AstraLux does not reveal a wide companion. The small elongation of the PSF is compatible 
with the predicted atmospheric dispersion effects, and cannot be taken as an indicator 
of an unresolved companion. The ability to image companions with AstraLux is highly 
dependent on the magnitude difference between the two stars, and GM Cep was always imaged 
during high magnitude phases, so the identification of companions 
at 100-200 AU would not be possible even if the two stars have a similar mass. 

\begin{figure}
\plotone{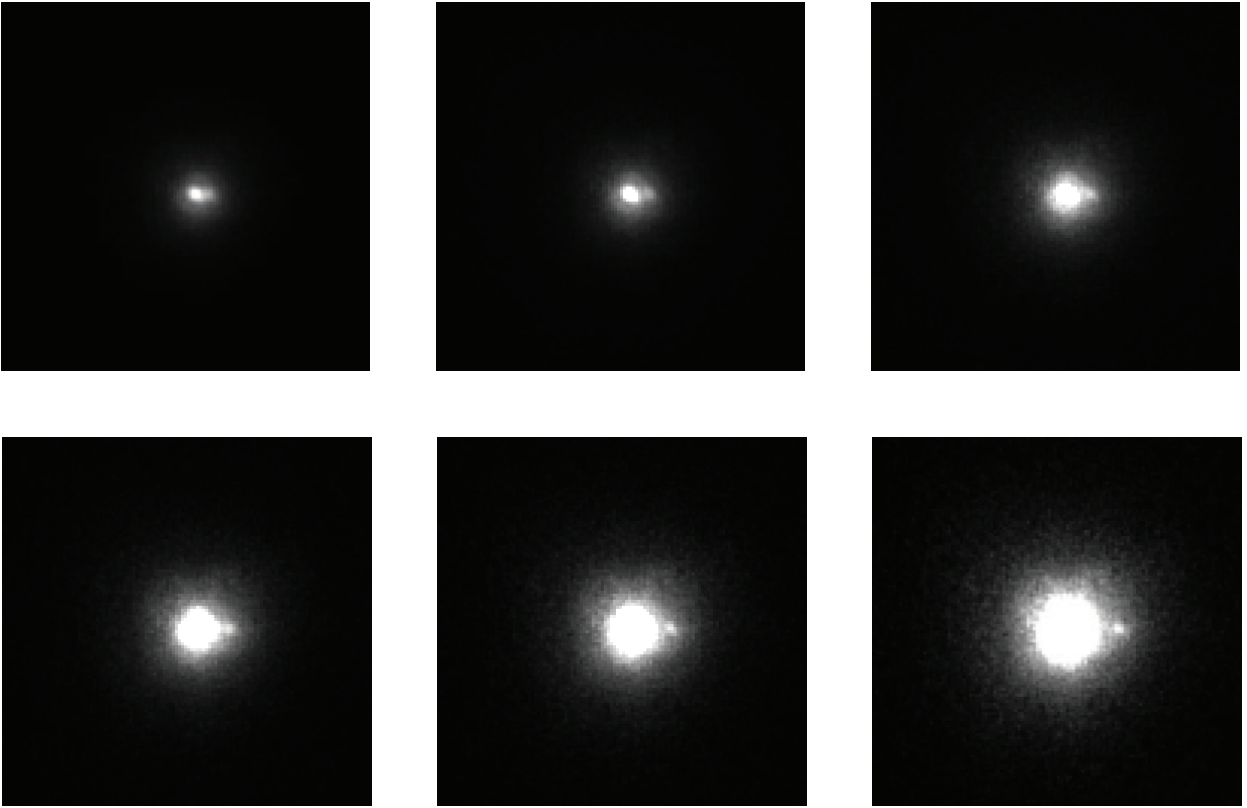}
\caption{Simulations of different binary cases as they would be imaged with AstraLux, 
for different separations and magnitude differences. From top to bottom, and 
left to right: separation 100 mas (90 AU), $\Delta$(mag)=1.20 mag; 
separation 150 mas (135 AU),$\Delta$(mag)=1.92 mag; separation 200 mas
(180 AU), $\Delta$mag)=2.20 mag; separation 250 mas (225 AU), 
$\Delta$(mag)=2.50 mag; separation 300 mas (270 AU), $\Delta$(mag)=3.00 mag; 
separation 400 mas (360 AU),$\Delta$(mag)=3.50 mag. 
All images are displayed on a linear scale up to saturation. The field
of view is 3''$\times$3''. 
\label{simulations-fig}}
\end{figure}

In order to constrain the separation and magnitude difference of a possible companion, we 
analyzed the maximum brightness differences of a hypothetical companions at different 
angular separations from GM Cep for the November 2006 AstraLux data, which has the best 
Strehl ratio. We adopted a 5$\sigma$ peak detection over the background noise as visibility 
criterion. The absence of static aberrations and asymmetries in the AstraLux data beyond the 
first diffraction ring allow to determine the intensity standard deviation in concentric 
annuli around GM Cep, resulting in robust estimates of the detection limits 
(Table \ref{binary-table}). Binary simulations were constructed by adding a scaled copy of 
the real observations to itself with the given magnitude differences and separations 
(Figure \ref{simulations-fig}). Since not only the stellar PSF is added, but also the noise, 
this method increases the background noise especially at small magnitude differences, leading 
to more pessimistic estimates of the achievable detection limits. The observed PSF is in 
good agreement with a theoretical simulation which includes atmospheric dispersion effects, 
so there is no evidence for a bright companion at angular separations between 50-100 mas
(45-90 AU). We conclude that GM Cep does not have a companion with a magnitude 
difference $\le$1 mag at $\approx$100 AU, and none farther than 200 AU with a magnitude
difference $\le$2 mag. However, we cannot rule out the presence of a close G- or K-type 
companion with normal luminosity. We will obtain further AstraLux observation of GM Cep near
its minimum in parallel to our photometric monitoring.
 
\section{Discussion: Potential Variability Mechanisms in GM Cep\label{outburst}}

There are different mechanisms responsible for the magnitude variations of 
pre-main sequence stars. Herbst et al. (1994) classifies them in Type I, 
due to rotation of a star with cool spots, Type II, due to rotation of a 
star with hot spots (related to accretion), and Type III, 
also called UX-or variables. Young, accreting, solar-type stars experience 
Type I and II variations of the order of few 0.01-0.5 mag in a few days 
(Bertout 1989; Brice\~{n}o et al. 2001; Eiroa et al. 2002). The irregular 
UX-or variations,  with more complex timescales of days/weeks combined with 
few-years cycles, do not usually exceed $\Delta$mag$\sim$1 mag, and occur 
typically in nearly edge-on HAeBe stars (Grinin et al. 2000), although other 
authors claim that stars as early as K0 can experience Type III variations
(Herbst \& Shevchenko 1999) and that the viewing angle can be as low as 45
degrees (Natta \& Whitney 2000). The mechanisms triggering the UX-or variations
are the most controversial, and they have been described as obscuration by shells 
of matter close to the star (Shevchenko et al. 1993; Natta et al. 1997; 
Grinin et al. 2000; Tambovtseva et al. 2001) or as thermal instabilities similar to
those triggering FU-or outbursts (Herbst \& Shevchenko 1999; Natta et al. 2000).
Strong variations ($\Delta$mag$\sim$2-5 mag) due to enhanced accretion episodes 
are seen in FU-or or EX-or outbursts, which are recurrent at least in the case 
of EX-or objects (Hartmann \& Kenyon 1996; Lehman et al. 1995; Herbig et al. 2001). 
These extreme accretors tend to be very fast rotators (Herbig et al. 2003; although 
the fast rotation detected may be actually the rotation of the very hot inner 
disk), to have IR fluxes that deviate from typical disk models (Hartmann \& Kenyon 1996), 
and to have in some cases companions at 100-300 AU (Reipurth \& Aspin 2004). FU-ors 
and probably EX-ors could be either a normal phase within the very early evolution 
and disk formation in TTS (Clarke et al. 1990; Hartmann \& Kenyon 1996, among others), 
or a special type of systems (certain binary or multiple systems; 
Herbig et al. 2003; Quanz et al. 2006). The long-time evolution
of these objects is not known, so the study of relatively old variables within
these classes could help to understand their nature.

The lightcurve of GM Cep is not purely consistent with any known FU-or, EX-or 
nor UX-or lightcurves, although these objects are largely variable within their 
classes. The magnitude changes ($\Delta$mag$\sim$2-2.5 mag) and multiple maxima 
are consistent with an EX-or or UX-or type variable. The high IR flux is another 
characteristic of EX-or/FU-or systems, due to their high luminosity and/or to 
the presence of envelopes. The H$\alpha$ and [N II] variability suggest changes 
in the accretion rates and in the strength of the winds, and although the 
H$\beta$ absorption is typical of UX-ors, the blueshifted absorption in H$\alpha$ 
is very rare in this class. The magnitude changes are not consistent with variable 
obscuration alone considering a standard galactic extinction law. Nevertheless,
if the extinction were produced by material in the circumstellar disk/shell
of GM Cep, containing reprocessed dust and large grains, the color changes during
high extinction episodes could be different. The type of obscuration suggested for UX-or
variables (Shevchenko et al. 1993; Natta et al. 1997) results in bluer colors at low 
magnitude, due to obscuration of the red disk and scattering (Grinin et al. 2001), which 
have not been seen so far in GM Cep. Moreover, pure extinction episodes cannot explain the 
high luminosity of GM Cep unless the spectral type and/or the luminosity class are 
changed beyond reasonable limits. Extinction variations alone could account 
only for up to $\sim$1.5 mags in V and cannot produce the observed changes in color.
If variable extinction were produced by the disk material in the vicinity 
of the star, the system should be close to edge-on or should have an anomalous 
distribution of circumstellar matter. Given the observed V$sini$, the angle of view must 
be larger than $\sim$15 degrees to prevent breakup, and the near-IR fluxes and 
the P-Cygni H$\alpha$ profile suggest an intermediate angle. If the excess 
mid-IR emission were due to an envelope, the strong winds could contribute to clean the 
polar gaps, reducing the extinction. GM Cep could be then one of the oldest EX-or variables 
known, since these phenomena seem to occur preferentially at ages $\sim$1 Myr. According 
to the picture developed by Clarke \& Syer (1996), the presence of a planetary or 
stellar companion in the disk of a Class II object at few AU distances can trigger 
thermal instability and produce outbursts with amplitude and frequency depending on 
the mass of the companion (Lodato \& Clarke 2004). Given the ages of Tr 37, it is 
most likely that GM Cep is a special object, probably a binary or multiple system, 
rather than a typical very young TTS.

In order to understand the variations of GM Cep, we compare it to similar cases:

The activity of the EX-ors EX Lupi and DR Tau resembles strongly that of GM Cep. 
EX Lupi suffers periods of quiescence followed by strong variability with
2-3 mag amplitude and changes in its emission line spectrum (Lehmann et al. 1995). 
Like GM Cep, it is presumably older than other EX-ors (Herbig 2007). The K5-K7 
eruptive star DR Tau is known to have irregular variability episodes since 1900, with 
amplitudes $\sim$1.5 to $\sim$3-5 mag. Its variable and strong stellar winds,
large UV excess and veiling suggest high and variable accretion rates (Chavarria 1979;
Mora et al. 2001). DR Tau can remain relatively stable at both the low-magnitude and 
high-magnitude phases during years/decades, as was seen by Suyarkova (1975) for GM Cep. 
The variability of DR Tau cannot be explained by hot and cold spots and variable obscuration
alone (Eiroa et al. 2002). Like for GM Cep, the mass of the disk of DR Tau 
($\sim$0.03 M$_\odot$; Beckwith et al. 1990) seems to be too small for its accretion rate 
if we assume the small typical grain sizes.

Suyarkova (1975) classified GM Cep as a RW Aur-type variable. The ``extreme CTTS'' RW Aur 
shares with GM Cep the strong and variable P-Cygni H$\alpha$ profile, a powerful disk, 
a large accretion rate, and a strong double-peaked O I emission lines at 8446 \AA, with 
velocity shifts $\pm$100 km/s (Ghez et al. 1993; Alencar et al. 2005). It is a triple 
system and the prototype of the RW Aur-class of irregular variable stars (Hoffmeister 1957),
characterized for rapid magnitude variations with $\Delta$mag$\sim$1-4 mag and a typical 
spectral type around G5.  The very small millimeter disk mass of RW Aur ($\sim$0.0003 M$_\odot$;  
Osterloh \& Beckwith 1995) suggests a timescale for disk removal of the order 
of centuries to millenia, too short unless we assume brief and transient accretion episodes 
(Cabrit 2005) and/or important grain growth to cm sizes (Rodmann et al. 2006). The accretion 
seems to be non-axisymmetric, as expected from interactions with a close companion
(Petrov et al. 2001).

The 1 Myr-old star GW Ori is one of the most massive TTS known (M=2.5 M$_\odot$, R=5.6 R$_\odot$; 
Mathieu et al. 1991), resembling strongly GM Cep. With an accretion rate $\sim$10$^{-6}$M$_\odot$/yr, 
it is a very fast rotator (V$sini$=43 km/s; Bouvier et al. 1986), and shows variability up to 1 mag
in JHK (Samus et al. 2004). It is a binary (maybe triple) system, with a $\sim$G5 primary with 
luminosity 26 L$_\odot$, similar to GM Cep and abnormally high for its spectral type (Mathieu et al. 1991).  
It is also  has a very strong IR excess, and is one of the most luminous TTS at millimeter wavelengths 
(Mathieu et al. 1991, 1995), suggesting a disk (or disk+shell) mass $\sim$1.5M$_\odot$, being the only 
one in its class able to sustain high accretion rates over several Myr. 
 
The K3 star CW Tau shares with GM Cep the rapid rotation (V$sini$=28 km/s; 
Muzerolle  et al. 1998), the H$\alpha$ P-Cygni profile and a deep, and a broad OI absorption 
at 7773\AA. It shows magnitude variations up to 2 magnitudes (General Catalog of Variable Stars,
Kukarkin et al. 1971), and has a strong outflow (Hirth et al. 1994; G\'{o}mez de Castro 1993) 
and strong [N II] forbidden line emission. As for GM Cep, the mass of the disk is not specially high
despite the signs of strong accretion ($<$0.02 M$_\odot$; Beckwith et al. 1990).

The outburst of IRAS 05436-0007/V1647 (Eisl\"{o}ffel \& Mundt 1997) revealed the 
bright McNeil Nebula in 2004 (McNeil et al. 2004). This very extincted (A$_V$$\sim$13 mag) 
object shows a flat mid-IR spectrum (in contrast to typical EX-or and similar objects, 
which have TTS-looking IR SEDs). It has been considered as a Class 0 embedded object 
(Lis et al. 1999), but may be some type of EX-or or FU-or Class II object, surrounded by 
a massive circumstellar envelope (M$\sim$0.5 M$_\odot$; \'{A}brah\'{a}m et al. 2004), 
with a 5.6 L$_\odot$ luminosity, typical of TTS, or an intermediate case between EX-or 
and FU-or objects (Fedele et al. 2007). Its emission line spectrum at optical wavelengths 
is similar to the spectrum of GM Cep.

KH 15D is a pre-main sequence binary system with a precessing disk or ring, seen
edge on, so one of the components is permanently occulted by the disk, whereas
the other is visible only during half of the period (Hamilton et al. 2005). 
The color changes are affected by the eclipses and scattering. 
Although such special configurations allow much large amplitudes and atypical color
variations, a similar scenario fails to explain the high luminosity of GM Cep without 
changing its spectral type and luminosity class or introducing additional mechanims (i.e., 
hot shells) to reproduce the optical absorption lines. 

\section{Conclusions \label{conclu}}

GM Cep is an extremely variable late G star in the 4 Myr-old cluster Tr 37, 
resembling younger EX-or objects. Its complex variability in color 
and magnitude is not consistent with a single typical variability 
mechanism in TTS (hot/cold spots, variable obscuration, 
or changes in the accretion rate). The amplitude of the variations, the
high accretion rate, the luminous mid-IR disk, and the high stellar 
luminosity suggest variable accretion to be the stronger contributor, maybe
mixed with variable extinction by circumstellar material (similar to
UX-or variables), and some minor influence of cold spots and scattering.
Increased accretion episodes are thought to produce the strong, irregular
variations in young EX-or and FU-or objects, typically still surrounded  by
massive infalling envelopes. Nevertheless, GM Cep has a medium-mass disk and a small
or inexistent envelope, and belongs to a cluster where strong disk evolution
is ubiquitous. Large changes in the accretion  rates can result
in changes in the stellar and disk structure, and conversely,
the disruption of the disk structure (for instance, by companions, or
via gravitational instability if the disk is very massive and compact),
can increase the accretion rate. The presence of companions 
(stellar, substellar or planetary) is a plausible mechanism to produce
disk instabilities in a relatively old star like GM Cep. Simultaneous 
multiwavelength observations, including IRS spectra scheduled for 2007, 
providing coverage in the 5-35$\mu$m range, and detailed 
monitoring in the coming years will help us to reveal the nature of the 
variability in GM Cep. Sub-millimetre observations in the future should 
be used to determine the SED slope and constrain the size of the grains.
Old but extremely accreting stars are thus a key to understand the processes 
of disk accretion and evolution and the consequences for planet formation.

\acknowledgements
We want to thank the Calar Alto Observatory for 
allocation of Director's Discretionary Time to this programme.
We thank L. A. Hillenbrand for kindly providing the HIRES spectrum
of GM Cep taken in 2001, J.A. Acosta-Pulido for providing us with his 'caindr'
package, and P. D'Alessio, for providing the tools for modelling the disk. 
P.\'A. acknowledges the support of the Hungarian Research Fund No. K62304. 
This publication makes use of data products from the Two Micron All Sky Survey, which is 
a joint project of the University of Massachusetts and the Infrared Processing and Analysis 
Center/California Institute of Technology, funded by the National Aeronautics and Space 
Administration and the National Science Foundation. This research has made use of the 
SIMBAD database, operated at CDS, Strasbourg, France, of the SuperCOSMOS Sky Survey,
provided by WFAU, Institute for Astronomy, Royal Observatory, Blackford Hill, 
Edinburgh, and of the Digitized Sky Surveys, produced at the Space Telescope Science 
Institute under U.S. Government grant NAG W-2166. We thank as well K. Tristram, 
S. Birkmann, B. Braunecker, and J. Stegmaier for their help at the 70 cm
K\"{o}nigstuhl telescope, and S. Pedraz-Marcos for his help with the Calar Alto
observations. We thank O. Fischer, J. Greiner, P. Kroll, D. Fedele,  C. Eiroa, 
M. Wiedner, H. Beuther,  V. Roccatagliata, and B. Conn for their comments and suggestions.
We finally thank the anonymous referee for his/her comments and suggestions to improve
the quality of this paper.

\clearpage

\input tab1.tex

\clearpage

\input tab2.tex

\input tab3.tex

\clearpage
\pagestyle{empty}
\input tab4.tex

\clearpage
\pagestyle{plaintop}
\input tab5.tex

\input tab6.tex

\input tab7.tex

\input tab8.tex

\clearpage

\end{document}

%% file: tab1.tex
\begin{deluxetable}{llcl}
\tabletypesize{\scriptsize}
%\rotate
\tablenum{1}
\tablecolumns{4} 
\tablewidth{0pc} 
\tablecaption{Summary of Observations on GM Cep\label{obs-table}} 
\tablehead{
 \colhead{Date/Epoch} & \colhead{Observatory/Telescope}  & \colhead{Filter/Mode} & \colhead{References}} 
\startdata
Before 1939 & Sonneberg Observatory & Visual Photographic Plates &  Morgenroth (1939) \\
1952.556$^a$ & Palomar & R63F &  USNO B1.0, Monet et al. (2003) \\
1955.936 &  Palomar  & GG395-B$_j$, R63F &  USNO A2.0, Monet et al. (1998) \\
1955.941 &  Palomar  & R63F &  SuperCosmos Catalog, Poss-I Red \\
1965-10-02 & Konkoly Observatory & V & Kun(1986) \\
1978.1: &  Palomar  & I9 &  USNO B1.0, Monet et al. (2003) \\
1983.5 & IRAS & 12-25-60-100$\mu$m &  IRAS (from Vizier) \\ 
1985-10-20 & Konkoly Observatory & V & Kun(1986) \\
1986-09-18 & Konkoly Observatory & V & Kun(1986) \\
1989.757 &  Palomar  & GG395-B$_j$  & USNO B1.0, Monet et al. (2003) \\
1990.763 &  Palomar  & R61F & SuperCosmos Catalog, Poss-II Red \\
1990.789: &  Palomar  & GG395-B$_j$ &  GSC 2.2 Catalogue \\
1990.789$^b$ &  Palomar  & R59F &  USNO B1.0, Monet et al. (2003) \\
1991.595$^c$ &  Palomar  & GG395-B$_j$ &  USNO B1.0, Monet et al. (2003) \\
1991.599 &  Palomar  & GG395-B$_j$ &  SuperCosmos Catalog, Poss-II Blue\\
1993.552 &  Palomar  & R63F &   GSC 2.2 Catalogue \\
1994.44 &  Palomar  & I9 &  SuperCosmos Catalog, Poss-II IR \\
1996-1998 & MSX6C & ACDE &  MSX6C, Egan et al. (2003) \\
2000-09-03 & FLWO 1.2m & R$_c$I$_c$ & Paper I \\
2000-09-04 & FLWO 1.2m & VR$_c$I$_c$ &  Paper I \\
2000-09-05 & FLWO 1.2m & R$_c$I$_c$ & Paper I \\
2000-09-06 & FLWO 1.2m & R$_c$I$_c$ &  Paper I \\
2000-09-07 & FLWO 1.2m & R$_c$I$_c$ &  Paper I \\
2000-09-08 & FLWO 1.2m & R$_c$I$_c$ &  Paper I \\
2000-09-09 & FLWO 1.2m & R$_c$I$_c$ &  Paper I \\
2000-09-21 & FLWO 0.9m 2MASS & JHK &    2MASS \\
2001-06-30 & Keck I/HIRES & High-res. spectroscopy  &  This work \\
2001-07-11 & FLWO 1.5m/FAST & Low-res. spectroscopy & Paper I \\
2001-09-21 & FLWO 1.2m & R$_c$I$_c$ & Paper I \\
2001-09-22 & FLWO 1.2m & R$_c$I$_c$ & Paper I \\
2001-09-23 & FLWO 1.2m & R$_c$I$_c$ & Paper I \\
2001-09-25 & FLWO 1.2m & R$_c$ &  Paper I \\
2002-09-03 & FLWO 1.2m & U &  Paper I \\
2003-10-20 & FLWO 1.2m & U &  Paper I \\
2003-12-20 & IRAC/Spitzer & 3.6, 4.5, 5.8, 8.0 $\mu$m &   Paper III \\
2004-06-23 & MIPS/Spitzer & 23.9, 70.0 $\mu$m & Paper III \\
2004-12-01 & Hectochelle/MMT & H$\alpha$ spectroscopy & Paper IV \\
2006-06-02/03 &  IRAM/MAMBO I & 1.3mm continuum &  This work \\
2006-06-03/05 &  IRAM/Heterodyne & $^{12}$CO(1-0), $^{12}$CO(2-1) &  This work \\
2006-07-07 & Calar Alto 2.2m/AstraLux & z' &  This work \\
2006-07-03 & K\"{o}nigstuhl 70 cm	& UBVRI	&   This work\\
2006-10-18 & K\"{o}nigstuhl 70 cm	& UBVRI	&   This work \\
2006-11-10 & Calar Alto 2.2m/AstraLux & z' &  This work \\
2006-12-15 & K\"{o}nigstuhl 70 cm	& UBVRI	&   This work \\
2006-12-29 & Calar Alto 2.2m/CAFOS & VRI	&  DDT program, this work \\
2006-12-31 & Calar Alto 2.2m/CAFOS & VRI	&  DDT program, this work \\
2007-01-02 & Calar Alto 2.2m/CAFOS & VRI	&  DDT program, this work \\
2007-01-18 & Calar Alto 2.2m/CAFOS & VRI	&  DDT program, this work \\
2007-04-27 & Calar Alto 2.2m/CAFOS & Low-res. spectroscopy	&  DDT program, this work \\
2007-05-11 & Calar Alto 2.2m/AstraLux & i'z' &  This work \\
2007-06-02 & Carlos S\'{a}nchez/Teide & JHK & This work \\
2007-06-03 & Carlos S\'{a}nchez/Teide & JHK & This work \\
2007-06-04 & Carlos S\'{a}nchez/Teide & JHK & This work \\
2007-06-05 & Carlos S\'{a}nchez/Teide & JHK & This work \\
2007-06-08 & Carlos S\'{a}nchez/Teide & JHK & This work \\
2007-06-08 & K\"{o}nigstuhl 70 cm	& UBVRI	&  This work \\
2007-06-24 & Calar Alto 2.2m/AstraLux & z' &  This work \\
2007-06-25 & Calar Alto 2.2m/AstraLux & z' &  This work \\
\enddata
\tablecomments{Summary of the data available on GM Cep, including our
own observations and the data from the literature.
1952.556$^a$: The epoch is uncertain, either 1952.556 or 1955.936.
1990.789$^b$: The epoch is uncertain, can be 1990.789, 1991.600, or 1993.070.
1991.595$^c$: The epoch is uncertain, either 1991.595 or 1991.603.}
\end{deluxetable}

%% file: tab2.tex
\begin{deluxetable}{lccccl}
\tabletypesize{\scriptsize}
%\rotate
\tablenum{2}
\tablecolumns{6} 
\tablewidth{0pc} 
\tablecaption{Summary of AstraLux Observations of GM Cep\label{astralux-table}} 
\tablehead{
 \colhead{Date} & \colhead{Filter}  & \colhead{T$_{int}$} & \colhead{Seeing} & \colhead{Strehl} & \colhead{Comments}} 
\startdata
2006-07-07    &      z'    &    3000x100ms   &   1.10   &     0.09 & \\
2006-11-10    &      z'    &    15000x25ms    &   0.60    &    0.14  & Atmospheric dispersion \\
2007-05-11    &      i'    &    10000x30ms   &   0.80    &    0.08  & \\
2007-05-11    &      z'    &    10000x30ms   &   0.80    &    0.10   & \\
2007-06-24    &      z'    &   15000x30ms    &  0.70    &    0.14  & \\
2007-06-25    &      z'    &   15000x30ms    &  0.85    &    0.12  &  \\
\enddata
\tablecomments{Summary of observations done with AstraLux. See text.}
\end{deluxetable}

%% file: tab3.tex
\begin{deluxetable}{lcccl}
\tabletypesize{\scriptsize}
%\rotate
\tablenum{3}
\tablecolumns{5} 
\tablewidth{0pc} 
\tablecaption{Near-Infrared Observations \label{JHK-table}} 
\tablehead{
\colhead{Epoch} & \colhead{ J}  & \colhead{H}  & \colhead{K} & \colhead{References}}
\startdata
22-09-2000  &    10.279$\pm$0.026 &    9.329$\pm$0.029  &  8.593$\pm$0.020 & 2MASS \\
02-06-2007  &    10.37$\pm$0.17   &    9.32$\pm$0.23    &  8.76$\pm$0.13 & This work \\
03-06-2007  &    10.52$\pm$0.07 &    9.36$\pm$0.05  &  8.48$\pm$0.04 & This work \\
04-06-2007  &    10.24$\pm$0.04 &    9.35$\pm$0.03  &  8.71$\pm$0.08 & This work \\
05-06-2007  &    10.21$\pm$0.04 &    9.26$\pm$0.03  &  8.64$\pm$0.03 & This work \\
08-06-2007  &    10.31$\pm$0.10 &    9.30$\pm$0.05  &  8.45$\pm$0.04 & This work \\
\enddata
\tablecomments{JHK magnitudes from 2MASS and from the observations at
the Carlos S\'{a}nchez telescope.}
\end{deluxetable}

%% file: tab4.tex
\begin{deluxetable}{lccccccccccccl}
\tabletypesize{\scriptsize}
\rotate
\tablenum{4}
\tablecolumns{14} 
\tablewidth{0pc} 
\tablecaption{Optical Data \label{opt-table}} 
\tablehead{
 \colhead{Epoch} &\colhead{U} & \colhead{B} &\colhead{V} & \colhead{R$_c$}  & \colhead{I$_c$}
  & \colhead{R$_J$}  & \colhead{I$_J$}  &\colhead{GG395-B$_j$}& \colhead{R59F} & \colhead{R61F} & \colhead{R63F}  &\colhead{I9}  &\colhead{Refs.} } 
\startdata
1952.556$^a$ & ---  & ---   &   --- &  12.92$\pm$0.1*   & ---  &  ---  & ---  & ---  &  --- & ---  & 12.84  & ---  & (1)  \\
1955.936 & ---  &  ---  &  ---  & 14.48$\pm$0.1*   & ---  &  ---  & ---  & 15.4($\pm$0.1)  &   &   & 14.4($\pm$0.1)  & ---  &  (2) \\
1955.941 & ---  & ---   &  ---  & 14.58$\pm$0.1*    & ---  &  ---  &  --- & ---  & ---  &  --- & 14.502  & ---  &   (3)\\
1978.1: & ---  & ---   & ---   &  ---   & ---  &  ---  & ---  & ---  & ---  & ---  & ---  & 11.77  & (1) \\
1965.753 & ---  & ---   & 12.87$\pm$0.1    & ---  &  ---  &  --- & ---  & ---  &  --- & --- & ---  & ---  &   (4)\\
1985.803 & ---  & 17.31:   & 14.30$\pm$0.1    & ---  &  ---  &  --- & ---  & ---  &  --- & --- & ---  & ---  &   (4)\\
1986.715 & ---  & ---   & 15.02$\pm$0.1    & ---  &  ---  &  --- & ---  & ---  &  --- & --- & ---  & ---  &   (4)\\
1989.575  & ---  & ---   & ---   &  ---   & ---  & ---   &  --- &  14.74 & ---  &   &  --- &  --- &  (1) \\
1990.763 & ---  &  ---  &  ---  &  ---   & ---  &  ---  & ---  & ---  & ---  & 13.993  &  --- & ---  &  (5) \\
1990.789: & ---  &  ---  & ---   &  ---   & ---  & ---   & ---  &  14.22 & ---  & ---  &  --- & ---  &  (6) \\
1990.789: & ---  &  ---  &  ---  &  ---   & ---  &  ---  & ---  &  14.52 & ---  &  --- &  --- & ---  &  (6) \\
1990.789$^b$ & ---  &  ---  & ---   &  14.05$\pm$0.1*   & ---  &  ---  & ---  &  --- & 14.08  & ---  & ---  & ---  &  (1) \\
1991.595$^c$ & ---  &  ---  &  ---  &  ---   & ---  & ---   & ---  &  14.44 &   & ---  & ---  & ---  &  (1) \\
1991.599 & ---  & ---   & ---   & ---    & ---  & ---   & ---  & 14.225  & ---  & ---  & ---  & ---  &  (7)\\
1993.552 & ---  &  ---  & ---   &  13.66$\pm$0.1*   & ---  &  ---  &  --- &  --- & ---  & ---  & 13.58  & ---  &  (6) \\
1994.44 & ---  & ---   & ---   & ---    & ---  & ---   & ---  & ---  & ---  & ---  & ---  & 12.995  &   (8) \\
2000.674 &  --- & ---   &  --- & 12.971$\pm$0.002   & 12.141$\pm$0.002  & ---  & ---  &  --- & ---  & ---  &  --- &  --- &    (9) \\
2000.677 & ---  & ---  & 13.915$\pm$0.002  & 12.952$\pm$0.002   &  12.141$\pm$0.002 &  12.52$\pm$0.07* & 11.7$\pm$0.2*  &  --- & ---  &  --- & ---  & ---  & (9) \\
2000.679 & ---  & ---   & ---  & 12.979$\pm$0.002   & 12.128$\pm$0.002  & ---  & ---  & ---  & ---  &  --- & ---  & ---  &    (9) \\
2000.684 & ---  & ---   & ---  & 12.988$\pm$0.002   & 12.213$\pm$0.002  & ---  & ---  & ---  & ---  &  --- & ---  & ---  &    (9) \\
2000.687 & ---  & ---   & ---  & 12.878$\pm$0.003   & 11.977$\pm$0.003  & ---  & ---  & ---  & ---  &  --- & ---  & ---  &    (9) \\
2000.690 & ---  & ---   & ---  & 12.894$\pm$0.004   & 12.030$\pm$0.002  & ---  & ---  & ---  & ---  &  --- & ---  & ---   &   (9) \\
2000.692 & ---  & ---   & ---  & 12.885$\pm$0.002   & 12.072$\pm$0.002  & ---  & ---  & ---  & ---  &  --- & ---  & ---  &   (9) \\
2001.723 & ---  & ---   & ---  & 12.702$\pm$0.003   & 11.846$\pm$0.002  & ---  & ---  & ---  & ---  &  --- & ---  & ---  &   (9) \\
2001.726 & ---  & ---   & ---  & 12.710$\pm$0.003   & 11.924$\pm$0.002  & ---  & ---  & ---  & ---  &  --- & ---  & ---  &   (9) \\
2001.729 & ---  & ---   & ---  & 12.753$\pm$0.003   & 11.857$\pm$0.003  & ---  & ---  & ---  & ---  &  --- & ---  & ---  &  (9) \\
2001.734 & --- & ---   & ---  & 12.715$\pm$0.003   &  --- & ---  &  --- & ---  & ---  &  --- & ---  & ---  &  (9) \\
2002.674 & 16.987$\pm$0.009  & ---   & ---  &  --- & ---  & ---  & ---  & ---  &  --- & ---  & ---  & ---  & (9) \\
2003.803 & 16.25$\pm$0.11  & ---   & ---  &  --- & ---  & ---  & ---  & ---  &  --- & ---  & ---  &  ---  & (9) \\
2006.504 & 15.04$\pm$0.02  & 14.51$\pm$0.01$^d$   & 13.173$\pm$0.005  &  12.36$\pm$0.04*  & 11.64$\pm$0.2*  & 12.024$\pm$0.011  & 11.20$\pm$0.01  & ---  & ---  &  --- & ---  & ---  & (10) \\
2006.797 & 15.64$\pm$0.03  & 15.02$\pm$0.04$^d$ & 13.613$\pm$0.004  &  12.76$\pm$0.03*  & 11.98$\pm$0.2*  & 12.405$\pm$0.007  & 11.525$\pm$0.008  & ---  & ---  &  --- & ---  & ---  & (10) \\
2006.956 & 16.02$\pm$0.05  & 15.33$\pm$0.01$^d$ & 13.783$\pm$0.004  &  12.88$\pm$0.03*  &  ---  & 12.467$\pm$0.007  & ---  & ---  & ---  &  --- & ---  & ---  & (10) \\
2006.995 &  --- & ---   & 14.802$\pm$0.007  & 13.96$\pm$0.04*   & 13.17$\pm$0.2*  & 13.60$\pm$0.02  &  12.714$\pm$0.005 & ---  & ---  &  --- & ---  & ---  & (10) \\
2006.998 &  --- & ---   & 14.729$\pm$0.004  &  --- &   &   & ---  & ---  &  --- & ---  & ---  & ---  &  (11) \\
2006.998 &  --- & ---   & 14.727$\pm$0.008  &  13.93$\pm$0.06*  & 13.15$\pm$0.2*  & 13.61$\pm$0.01    & 12.706$\pm$0.005 & ---  & ---  &  --- & ---  & ---  & (11) \\ 
2007.005 &  --- & ---   & 14.728$\pm$0.008  &  13.91$\pm$0.04*  & 13.19$\pm$0.2*  & 13.571$\pm$0.009  & 12.749$\pm$0.006 & ---  & ---  &  --- & ---  & ---  & (11) \\
2007.049 &  --- & ---   & 13.912$\pm$0.006  &  13.01$\pm$0.06*  &	12.02$\pm$0.2* & 12.61$\pm$0.01	   & 11.507$\pm$0.005   & ---  & ---  &  --- & ---  & ---  & (11) \\
2007.321 &  --- & ---   & 14.41$\pm$0.01  &  13.46$\pm$0.06*  &	--- & 13.00$\pm$0.01   & ---   & ---  & ---  &  --- & ---  & ---  & (12) \\
2007.435 &  --- & ---   & 13.26$\pm$0.02 & --- & --- & --- & ---   & ---  & ---  &  --- & ---  & ---  & (10) \\
2007.435 & 15.26$\pm$0.13  & 14.63$\pm$0.03  & 13.302$\pm$0.009  & 12.43$\pm$0.08*  & 11.73$\pm$0.2*  & 12.06$\pm$0.02 &  11.29$\pm$0.01 & ---  & ---  &  --- & ---  & ---  & (10) \\
2007.435 &  --- & ---   & 13.24$\pm$0.02 & --- & --- & --- & ---   & ---  & ---  &  --- & ---  & ---  & (10) \\
\enddata 
\tablecomments{Optical data available for GM Cep. Asterisk indicates that the magnitude has
been transformed into the given system, following the tranformations of Fernie (1983), Bessel (1979),
Bessell et al. (1986),Blair \& Gilmore (1982), Couch \& Newell (1980). Uncertain magnitudes
and epochs are marked by ``:''.
1952.556$^a$: The epoch is uncertain, either 1952.556 or 1955.936.
1990.789$^b$: The epoch is uncertain, can be 1990.789, 1991.600, or 1993.070.
1991.595$^c$: The epoch is uncertain, either 1991.595 or 1991.603.
The $^d$ index for the B band data from the
K\"{o}nigstuhl indicates that the zeropoint is uncertain (see Section \ref{data-optical}).
The references to the data catalog and/or telescope and instruments are: (1) = USNO B1.0, Monet et al.(2003); 
(2) = USNO-A2.0, Monet et al. (1998);(3) = SuperCosmos Catalog, Poss-I Red; 
(4) = Konkoly Observatory, Kun (1986); (5) = SuperCosmos Catalog, Poss-II Red; (6) = GSC 2.2 Catalogue;
(7) = SuperCosmos Catalog, Poss-II Blue; (8) = SuperCosmos Catalog, Poss-II IR; (9) = 4Shooter/48'' FLWO;
(10) = 70cm K\"{o}nigstuhl; (11)=CAFOS/2.2m Calar Alto; (12)=CAFOS/2.2m Calar Alto spectroscopy mode.}
\end{deluxetable}

%% file: tab5.tex
\begin{deluxetable}{lcccl}
\tabletypesize{\scriptsize}
%\rotate
\tablenum{5}
\tablecolumns{5} 
\tablewidth{0pc} 
\tablecaption{Emission and Absorption Lines\label{lines-table}} 
\tablehead{ \colhead{Line} & \colhead{$\lambda$ (\AA)}  & \colhead{EW (\AA)}  & \colhead{Epoch} & \colhead{Comments}  } 
\startdata
$[$N II$]$  & 6548 & -0.02   & HIRES 2001  & \\
''  & '' & -0.04   & Hecto. 2004 & \\
H$\alpha$ & 6563 & -6 & HIRES 2001 & Bluesh. abs. V$\sim$-80 km/s, B=-2.0\AA, R=-4.0\AA  \\
'' & '' & -10 & FAST 2001 & \\
'' & '' & -14 & Hectochelle 2004 & Bluesh. abs. V$\sim$-60 km/s, B=-3.5\AA, R=-10.5\AA  \\
'' & '' & -19 & CAFOS 2007 & \\
$[$N II$]$  & 6583 & -0.03   & HIRES 2001  & \\
''  & '' & -0.08   & Hecto. 2004 & \\
$[$S II$]$  & 6717 & --- & HIRES 2001 & \\
$[$S II$]$  & 6731 & -0.12 & HIRES 2001 & V$\sim$-100 km/s \\
Fe II   & 6516 & -0.06 & Hectochelle 2004 &   \\
Fe I     & 6495 & -0.04 & Hectochelle 2004 & \\ 
Li I    & 6708 &  0.16 & HIRES 2001 \\
''    & '' & ---   & FAST 2001  & Probably undetected because of veiling\\
''    & '' & ---   & CAFOS 2007 & \\
$[$O I$]$   & 6363 & --- & HIRES 2001 & May be masked by nearby absorption lines\\
O I 	& 7774 & 0.79 & HIRES 2001 & \\
'' 	& ''   & 1.2  & CAFOS 2007 & \\
O I     & 8446 & -0.52 & HIRES 2001 & V$\sim$ -140/+80 km/s, B=-0.32\AA, R=-0.20\AA \\
''      & '' & -0.2 & CAFOS 2007 &  \\
Ca II   & 8498 & -3.08 & HIRES 2001 &\\
''   & '' & -3.8  & CAFOS 2007 &\\
Ca II   & 8542 & -2.2  & CAFOS 2007 & \\
Ca II   & 8662 & -1.25 & HIRES 2001 & May be blended with H Pa 13 \\
''  & '' & -2.5  & CAFOS 2007 & May be blended with H Pa 13 \\
\enddata
\tablecomments{Emission and absorption lines that related to accretion and/or winds.
Negative and positive EW correspond to emission and absorption lines, respectively.
The CAFOS 2007 data has simultaneous photometry showing V=14.41$\pm$0.01 and R$_J$=13.00$\pm$0.01.
``B'' and ``R'' indicate the EW of the blue- and redshifted components, respectively.}
\end{deluxetable}

%% file: tab6.tex
\begin{deluxetable}{lccl}
\tabletypesize{\scriptsize}
%\rotate
\tablenum{6}
\tablecolumns{4} 
\tablewidth{0pc} 
\tablecaption{Mid-Infrared Fluxes \label{IR-table}} 
\tablehead{
\colhead{Epoch} & \colhead{ $\lambda$/Band ($\mu$m)}  & \colhead{Flux (Jy)}& \colhead{References}}
\startdata
1983.5 		& 12		& 0.659$\pm$0.046	& IRAS, computed by Vizier	\\
1983.5 		& 25		& 0.836$\pm$0.075	& IRAS, computed by Vizier	\\
1983.5 		& 60		& 1.44$\pm$0.20		& IRAS, computed by Vizier	\\
1983.5 		& 100		& $<$2.60		& IRAS, computed by Vizier	\\
1996-1998	& 8.28/A	& 0.622$\pm$0.031	& MSX6C, (Egan et al. 2003)	\\
1996-1998	& 12.13/C	& $<$0.629		& MSX6C, (Egan et al. 2003)	\\
1996-1998	& 14.65/D	& $<$0.482		& MSX6C, (Egan et al. 2003)	\\
1996-1998	& 21.0/E	& $<$1.37		& MSX6C, (Egan et al. 2003)	\\
2003.970	& 3.6		& 0.247$\pm$0.012	& IRAC/Spitzer (Paper III)	\\
2003.970	& 4.5		& 0.219$\pm$0.011	& IRAC/Spitzer (Paper III)	\\
2003.970	& 5.8		& 0.256$\pm$0.013	& IRAC/Spitzer (Paper III)	\\
2003.970	& 8.0		& 0.282$\pm$0.014	& IRAC/Spitzer (Paper III)	\\
2004.477	& 23.9		& 0.791$\pm$0.079	& MIPS/Spitzer (Paper III)	\\
2004.477	& 70.0		& 0.829$\pm$0.083	& MIPS/Spitzer (Paper III)	\\
\enddata
\tablecomments{Mid-IR fluxes from the literature.}
\end{deluxetable}

%% file: tab7.tex
\begin{deluxetable}{lccccc}
\tabletypesize{\scriptsize}
%\rotate
\tablenum{7}
\tablecolumns{6} 
\tablewidth{0pc} 
\tablecaption{Radial and Rotational Velocity\label{xcor-table}} 
\tablehead{
 \colhead{$<\lambda >$ (\AA)} & \colhead{Standard/Sp.Type}  & \colhead{R}
  & \colhead{CZ (km/s)}  & \colhead{V$sini$ (km/s)} & \colhead{Comments}} 
\startdata
6370 & HD222368/F7V & 3.4 & -22.1$\pm$8.1 & 49.9$\pm$11.5 & \\
6470 & HD222368/F7V & 10.9 & -23.9$\pm$2.7 & 41.2$\pm$3.5 & \\
6590 & HD222368/F7V & 1.3 & -19.7$\pm$12.7 & 44.2$\pm$19.4 & Few lines \\
6710 & HD222368/F7V & 4.0 & -30.7$\pm$5.8 & 43.3$\pm$8.7 & \\
6840 & HD122693/F8V & 1.6 & -27.2$\pm$12.0 & 52.5$\pm$20.4 & Few lines \\
7110 & HD222368/F7V & 5.1 & -8.0$\pm$3.9 & 49.1$\pm$8.1 & \\
7410 & HD222368/F7V & 6.7 & -25.6$\pm$4.3 & 47.7$\pm$6.2 & \\
7730 & HD122693/F8V & 3.5 & -26.3$\pm$7.0 & 50.4$\pm$11.3 & \\
7910 & HD222368/F7V & 4.7 & -14.9$\pm$4.7 & 41.0$\pm$7.2 & \\
8670 & HD126053/G1V & 2.2 & -25.4$\pm$9.9 & 42.7$\pm$13.5 & \\
8270 & HD222368/F7V & 1.6 & -13.5$\pm$9.5 & 40.0$\pm$15.6 & Few lines\\
8470 & HD122693/F8V & 1.4 & -29.0$\pm$10.7 & 38.0$\pm$15.7 & Few lines \\
8470 & HD126053/G1V & 1.4 & -28.1$\pm$12.0 & 46.0$\pm$19.0 & Few lines \\
\enddata
\tablecomments{Results of the cross-correlation for the different 
orders of the HIRES spectrum, including the standard used for comparison
and its spectral type, the R parameter, indicative of the goodness of the
correlation (Tonry \& Davis 1979; Hartmann et al. 1986; Kurtz et al. 1992),
the radial velocity CZ, and the rotational velocity V$sini$. Some parts of 
the spectra are affected by H$_2$O and O$_2$ atmospheric absorption bands, 
leading to small regions free of absorption, few lines to correlate, and
worse correlations (R$<$2). The orders centered in 6970 and 8090\AA\
are so strongly affected by atmospheric absorption features that we
could not cross-correlate them.}
\end{deluxetable}

%% file: tab8.tex
\begin{deluxetable}{lcc}
\tabletypesize{\scriptsize}
%\rotate
\tablenum{8}
\tablecolumns{3} 
\tablewidth{0pc} 
\tablecaption{Binary Detection Limits For AstraLux\label{binary-table}} 
\tablehead{
 \colhead{Distance (AU)} & \colhead{Separation ('')}  & \colhead{$\Delta$(mag)}} 
\startdata
60 & 0.07 & -0.5 \\
90 & 0.10 & -1.2 \\
100 & 0.11 & -1.8 \\
150 & 0.17 & -2.0 \\
200 & 0.22 & -2.6 \\
250 & 0.27 & -3.2 \\
300 & 0.33 & -3.5 \\
350 & 0.39 & -3.8 \\
400 & 0.44 & -3.9 \\
425 & 0.47 & -4.0 \\
450 & 0.50 & -4.1 \\
500 & 0.56 & -4.4 \\
600 & 0.67 & -4.6 \\
700 & 0.78 & -4.8 \\
800 & 0.89 & -5.0 \\
900 & 1.00 & -5.3 \\
1000 & 1.11 & -5.5 \\
1500 & 1.68 & -6.2 \\
2000 & 2.22 & -6.4 \\
2500 & 2.78 & -6.6 \\
3000 & 3.33 & -6.6 \\
3500 & 3.89 & -6.6 \\
\enddata
\tablecomments{Detection limits for binary companions with AstraLux. Based on a
5$\sigma$ detection over the background.}
\end{deluxetable}